\documentclass[usenatbib]{mn2e} 
\usepackage{graphicx,fixltx2e,amsmath,amssymb,amsfonts,latexsym,wasysym}

\def\chisq{\hbox{$\chi^2$}}
\def\chisqr{\hbox{$\chi^2_{\rm r}$}}
\def\msun{\hbox{${\rm M}_{\odot}$}}
\def\mjup{\hbox{${\rm M}_{\rm Jup}$}}
\def\mspy{\hbox{${\rm M}_{\odot}$\,yr$^{-1}$}}
\def\rsun{\hbox{${\rm R}_{\odot}$}}
\def\lsun{\hbox{${\rm L}_{\odot}$}}
\def\rcor{\hbox{$r_{\rm cor}$}}
\def\rmag{\hbox{$r_{\rm mag}$}}

\def\rstar{\hbox{$R_{\star}$}}
\def\lstar{\hbox{$L_{\star}$}}
\def\teff{\hbox{$T_{\rm eff}$}}
\def\logg{\hbox{$\log g$}}
\def\sn{\hbox{S/N}}
\def\vrad{\hbox{$v_{\rm rad}$}}

\def\kms{\hbox{km\,s$^{-1}$}}
\def\vsini{\hbox{$v \sin i$}}

\def\ptt{\hbox{$10^{-4} I_{\rm c}$}}

\def\arcsec{\hbox{$^{\prime\prime}$}}
\def\degr{\hbox{$^\circ$}}

\newcommand{\caii}{\hbox{Ca$\;${\sc ii}}}
\newcommand{\fei}{\hbox{Fe$\;${\sc i}}}

\newcommand{\hei}{\hbox{He$\;${\sc i}}}
\newcommand{\hal}{\hbox{H${\alpha}$}}
\newcommand{\hbe}{\hbox{H${\beta}$}}

\begin{document}

\title[Magnetometry of the cTTS GQ~Lup]{Magnetometry of the classical T~Tauri star GQ~Lup: non-stationary dynamos \& spin evolution of young Suns} 
\makeatletter

\def\newauthor{%
  \end{author@tabular}\par
  \begin{author@tabular}[t]{@{}l@{}}}
\makeatother
 
\author[J.-F.~Donati et al.]
{\vspace{1.7mm}
J.-F.~Donati$^1$\thanks{E-mail: 
jean-francois.donati@irap.omp.eu }, 
S.G.~Gregory$^2$, S.H.P.~Alencar$^3$, G.~Hussain$^4$, J.~Bouvier$^5$, \\  
\vspace{1.7mm}
{\hspace{-1.5mm}\LARGE\rm 
C.~Dougados$^5$, M.M.~Jardine$^6$, F.~M\'enard$^5$, M.M.~Romanova$^7$ \& the MaPP}\\ 
\vspace{1.7mm}
{\hspace{-1.5mm}\LARGE\rm
collaboration} \\
$^1$ IRAP--UMR 5277, CNRS \& Univ.\ de Toulouse, 14 Av.\ E.~Belin, F--31400 Toulouse, France \\
$^2$ California Institute of Technology, MC 249-17, Pasadena, CA 91125, USA \\ 
$^3$ Departamento de F\`{\i}sica -- ICEx -- UFMG, Av. Ant\^onio Carlos, 6627, 30270-901 Belo Horizonte, MG, Brazil \\ 
$^4$ ESO, Karl-Schwarzschild-Str.\ 2, D-85748 Garching, Germany \\ 
$^5$ IPAG--UMR 5274, CNRS \& Univ.\ J.~Fourier, 414 rue de la Piscine, F--38041 Grenoble, France \\ 
$^6$ School of Physics and Astronomy, Univ.\ of St~Andrews, St~Andrews, Scotland KY16 9SS, UK \\ 
$^7$ Department of Astronomy, Cornell University, Ithaca, NY 14853-6801, USA  
}

\date{2012 June, MNRAS in press}
\maketitle
 
\begin{abstract}  

We report here results of spectropolarimetric observations of the classical T~Tauri star (cTTS) 
GQ~Lup carried out with ESPaDOnS at the Canada-France-Hawaii Telescope (CFHT) in the framework 
of the `Magnetic Protostars and Planets' (MaPP) programme, and obtained at 2 different epochs 
(2009~July and 2011~June).  From these observations, we first infer that GQ~Lup has a photospheric 
temperature of $4,300\pm50$~K and a rotation period of $8.4\pm0.3$~d;  it implies that it is a 
$1.05\pm0.07$~\msun\ star viewed at an inclination of $\simeq$30\degr, with an age of 2--5~Myr,  
a radius of $1.7\pm0.2$~\rsun, and has just started to develop a radiative core.  

Large Zeeman signatures are clearly detected at all times, both in photospheric lines and in 
accretion-powered emission lines, probing longitudinal fields of up to 6~kG and hence making 
GQ~Lup the cTTS with the strongest large-scale fields known as of today.  Rotational modulation 
of Zeeman signatures, also detected both in photospheric and accretion proxies, is clearly different 
between our 2 runs;  we take this as further evidence that the large-scale fields of cTTSs are 
evolving with time and thus that they are produced by non-stationary dynamo processes.  

Using tomographic imaging, we reconstruct maps of the large-scale field, of the photospheric 
brightness and of the accretion-powered emission at the surface of GQ~Lup at both epochs.  
We find that the magnetic topology is mostly poloidal and axisymmetric with respect to the rotation 
axis of the star;  moreover, the octupolar component of the large-scale field (of polar strength 2.4 and 1.6~kG in 2009 
and 2011 respectively) dominates the dipolar component (of polar strength $\simeq$1~kG) by a factor of $\simeq$2, 
consistent with the fact that GQ~Lup is no longer fully-convective.  

GQ~Lup also features dominantly poleward magnetospheric accretion at both epochs.  The large-scale 
dipole component of GQ~Lup is however not strong enough to disrupt the surrounding accretion disc 
further than about half-way to the corotation radius (at which the Keplerian period of the disc material equals the 
stellar rotation period), suggesting that GQ~Lup should rapidly spin up like other similar 
partly-convective cTTSs.  

We finally report a 0.4~\kms\ RV change for GQ~Lup between 2009 and 2011, suggesting that  
a brown dwarf other than GQ~Lup~B may be orbiting GQ~Lup at a distance of only a few au's.  
\end{abstract}


\section{Introduction} 
\label{sec:int}

It is now well recognised that magnetic fields can significantly modify the life 
of stars and in particular their rotation rates.  Their impact is thought to be 
strongest throughout the formation stages, when stars and their planetary systems 
build up from the collapse of giant molecular clouds.  More specifically, fields are 
likely efficient at slowing down the cloud collapse, at inhibiting the subsequent 
fragmentation and at dissipating the cloud angular momentum through magnetic 
braking and the associated magnetised outflows / collimated jets \citep[e.g.,][for 
reviews]{Donati09, Andre09}.  At a later stage, the newly born protostars (called 
classical T~Tauri stars or cTTSs) are apparently capable of generating magnetic 
fields strong enough to disrupt the central regions of their accretion discs and 
to funnel some of the inner disc material onto the stellar surface, thereby 
drastically modifying the overall mass accretion process 
\citep[e.g.,][]{Bouvier07}.  

For some time, the strong magnetic fields of cTTSs could only be inferred through 
indirect proxies such as continuum or line emission throughout the whole 
electromagnetic spectrum, from X-rays to radio wavelengths.  Directly detected 
for the first time about 2 decades ago through the Zeeman broadening they induce 
on spectral lines observed in unpolarized light \citep[e.g.,][for a recent 
overview]{Johns07}, magnetic fields of cTTSs can now be characterized by various 
means.  In particular, their large-scale topologies - controlling how the fields 
couple the central protostars to the inner regions of their accretion discs and 
thereby how disc material is being accreted - can be thoroughly investigated 
thanks to the advent of sensitive high-resolution spectropolarimeters dedicated to
the study of stellar magnetic fields \citep[e.g.,][]{Donati03, Donati09}.  By 
measuring circularly-polarised Zeeman signatures of cTTSs and by monitoring their 
rotational modulation, one can reconstruct the parent large-scale magnetic 
topologies, thus offering the possibility of studying magnetospheric accretion 
processes in a much more quantitative way.  

The international MaPP (Magnetic Protostars and Planets) project was designed mostly 
for this purpose.  The first main goal of MaPP is to investigate (through a first 
survey of $\simeq15$ targets, with some of them observed at several epochs) how the 
large-scale magnetic topologies of cTTSs depend on key stellar parameters such as 
mass, age, rotation and accretion rate \citep[e.g.,][]{Donati10b, Donati11b}.  
The second main goal of MaPP is to provide an improved theoretical description, 
using both analytical modelling and numerical simulations, of how magnetic fields 
of cTTSs are generated and how they modify mass accretion processes \citep[see, 
e.g.,][]{Gregory10, Romanova11}, and more generally how critically they impact the 
formation of low-mass stars.  A total of 690~hr of time was allocated for MaPP on 
the 3.6~m Canada-France-Hawaii Telescope (CFHT) over a timescale of 9 semesters 
(2008b-2012b).  Up to now magnetic Zeeman signatures were detected on all selected 
targets;  several major discoveries were achieved regarding the two science goals 
mentioned above, that we will recall below in the light of the new results 
presented here.  

This new study focusses on the cTTS GQ~Lup, whose mass is a fair match to that of 
the Sun and whose age is typical of cTTSs.  
Located near Lupus~1, in the Lupus star formation region \citep[$150\pm10$~pc away from
the Earth,][]{Wichmann99, Crawford00}, GQ~Lup recently
attracted a lot of attention following the discovery of its low-mass companion \citep[most
likely a brown dwarf, with a mass in the range 10--40~\mjup,][]{McElwain07,Lavigne09}
in the outer regions of its accretion disc \citep[at a distance of $\simeq$0.7\arcsec\ or
100~au,][]{Neuhauser05}.  Although the nature of this sub-stellar companion is still a
matter of speculation, it makes GQ~Lup an obvious target of study, to investigate the
properties of the central protostar on the one hand (and in particular the star-disc
interaction in which the large-scale magnetic field of the protostar plays a crucial role)
and to better understand how stellar / planetary systems form.  In this respect, efforts
at modelling the magnetic field of the protostar and associated activity are worthwhile.
The large-scale field is obviously a key parameter to unravel the physics of the
star-disc magnetospheric interaction, whereas the potential presence of closer
companions (e.g., that could explain the ejection on an outer orbit of the very distant
companion detected already) can only be revealed through high-precision radial velocity
(RV) measurements when an accurate description of the magnetic activity, and an
efficient way of filtering the associated RV jitter, become available.

We start this paper by describing the spectropolarimetric observations of GQ~Lup we collected 
at 2 different epochs and from which Zeeman signatures are clearly detected (Sec.~\ref{sec:obs}).  
Following a fresh re-determination of the main characteristics of this cTTS (Sec.~\ref{sec:gq}), 
we outline the rotational modulation and intrinsic long-term variability that we 
observe in the data (Sec.~\ref{sec:var}).  We then detail the modelling of these 
data with our magnetic imaging code (Sec.~\ref{sec:mod}), compare our new results 
with previous ones and outline how they improve our understanding 
of how magnetic fields impact the formation of Sun-like stars (Sec.~\ref{sec:dis}).

\section{Observations}
\label{sec:obs}

Spectropolarimetric observations of GQ~Lup were collected at two different epochs, first from
2009~July 01 to 14, then from 2011~June 08 to 23, using the high-resolution spectropolarimeter
ESPaDOnS at the Canada-France-Hawaii Telescope (CFHT).  ESPaDOnS collects stellar spectra
spanning the entire optical domain (from 370 to 1,000~nm) at a resolving power of 65,000
(i.e., resolved velocity element of 4.6~\kms), in either circular or linear polarisation
\citep{Donati03}.
In 2009 (resp.\ 2011), a total of 14 (resp.\ 12) circular polarisation spectra were collected
over a timespan of 14 (resp.\ 16) nights with as regular a time sampling as possible (about 1
spectrum per night) given weather conditions.
All polarisation spectra consist of 4 individual subexposures (each lasting 904~s and 853.5~s
in 2009 and 2011 respectively) taken in different polarimeter configurations to allow the
removal of all spurious polarisation signatures at first order.
All raw frames are processed as described in the previous papers of the series 
\citep[e.g.,][]{Donati10b, Donati11}, to which the reader is referred for more information.  
The peak signal-to-noise ratios (\sn, per 2.6~\kms\ velocity bin) achieved on the
collected spectra range between 60 and 250 depending on weather/seeing conditions
with a median value of about 200.
The full journal of observations is presented in Table~\ref{tab:logesp}.

\begin{table}
\caption[]{Journal of observations collected in 2009~July and 2011~June.
Each observation consists of a sequence of 4 subexposures (each lasting 904~s
and 853.5~s in 2009 and 2011 respectively).
Columns $1-4$ respectively list the UT date, the Heliocentric Julian Date (HJD) and
UT time (both at mid-exposure), and the peak signal to noise ratio (per 2.6~\kms\
velocity bin) of each observation.
Column 5 lists the rms noise level (relative to the unpolarized continuum level
$I_{\rm c}$ and per 1.8~\kms\ velocity bin) in the circular polarization profile
produced by Least-Squares Deconvolution (LSD), while column~6 indicates the
orbital/rotational cycle associated with each exposure (using the ephemeris given by
Eq.~\ref{eq:eph}).  }
\begin{tabular}{cccccc}
\hline
Date & HJD          & UT      &  \sn\  & $\sigma_{\rm LSD}$ & Cycle \\
2009 & (2,455,000+) & (h:m:s) &      &   (\ptt)  & (0+) \\
\hline
Jul 01 & 13.81216 & 07:25:13 & 140 & 3.3 & 0.097 \\
Jul 02 & 14.81245 & 07:25:45 & 180 & 2.4 & 0.216 \\
Jul 03 & 15.80196 & 07:10:45 & 200 & 2.2 & 0.334 \\
Jul 04 & 16.79606 & 07:02:22 & 180 & 2.5 & 0.452 \\
Jul 05 & 17.79593 & 07:02:17 & 200 & 2.2 & 0.571 \\
Jul 06 & 18.79716 & 07:04:11 & 180 & 2.4 & 0.690 \\
Jul 07 & 19.79736 & 07:04:34 & 120 & 4.1 & 0.809 \\
Jul 08 & 20.79778 & 07:05:18 & 170 & 2.7 & 0.928 \\
Jul 09 & 21.79511 & 07:01:34 & 120 & 3.9 & 1.047 \\
Jul 10 & 22.79485 & 07:01:18 & 190 & 2.3 & 1.166 \\
Jul 11 & 23.80029 & 07:09:15 &  60 & 8.8 & 1.286 \\
Jul 12 & 24.81958 & 07:37:09 & 190 & 2.4 & 1.407 \\
Jul 13 & 25.75895 & 06:09:58 & 210 & 2.1 & 1.519 \\
Jul 14 & 26.75935 & 06:10:39 & 190 & 2.5 & 1.638 \\
\hline
Date & HJD          & UT      &  \sn\  & $\sigma_{\rm LSD}$ & Cycle \\
2011  & (2,455,700+) & (h:m:s) &      &   (\ptt)  & (84+) \\
\hline
Jun 08 & 20.96469 & 11:02:45 & 140 & 3.3 & 0.281 \\
Jun 11 & 23.91064 & 09:45:08 & 240 & 1.7 & 0.632 \\
Jun 12 & 24.93835 & 10:25:06 & 240 & 1.8 & 0.755 \\
Jun 14 & 26.86920 & 08:45:42 & 180 & 2.5 & 0.984 \\
Jun 15 & 27.88564 & 09:09:27 & 240 & 1.8 & 1.105 \\
Jun 16 & 28.88107 & 09:02:57 & 190 & 2.3 & 1.224 \\
Jun 17 & 29.84680 & 08:13:41 & 190 & 2.2 & 1.339 \\
Jun 18 & 30.91269 & 09:48:40 & 220 & 2.0 & 1.466 \\
Jun 20 & 32.91118 & 09:46:40 & 250 & 1.8 & 1.704 \\
Jun 21 & 33.84406 & 08:10:07 & 250 & 1.8 & 1.815 \\
Jun 22 & 34.93027 & 10:14:21 & 240 & 1.9 & 1.944 \\
Jun 23 & 35.91838 & 09:57:20 & 210 & 2.2 & 2.062 \\
\hline
\end{tabular}
\label{tab:logesp}
\end{table}

As outlined in Sec.~\ref{sec:var}, we find that the rotation period of GQ~Lup is 
$8.4\pm0.3$~d, i.e., very similar to that derived by \citet{Broeg07}.
Rotational cycles $E$ of GQ~Lup are thus computed from Heliocentric Julian Dates (HJDs)
according to the following ephemeris:
\begin{equation}
\mbox{HJD} = 2455013.0 + 8.4 E.
\label{eq:eph}
\end{equation}
Coverage of the rotation cycle is reasonably dense and regular in 2009 and 2011, at
which $\simeq$1.6 and 2 complete cycles of GQ~Lup were monitored, offering us a
convenient way of disentangling intrinsic variability from rotational modulation in
the spectra (see Sec.~\ref{sec:var}).

\begin{figure}
\includegraphics[scale=0.35,angle=-90]{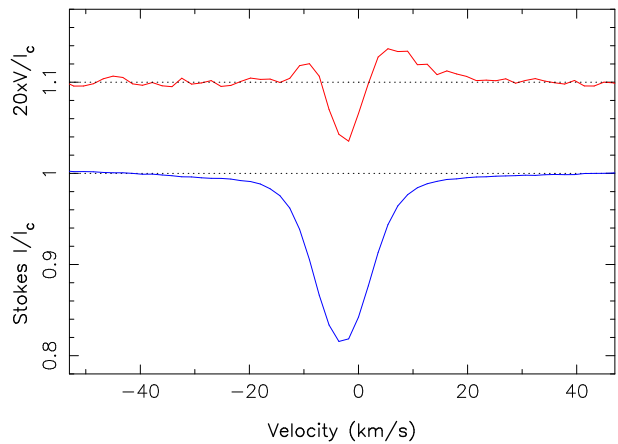}
\caption[]{LSD circularly-polarized (Stokes $V$) and unpolarized (Stokes $I$)
profiles of GQ~Lup (top/red, bottom/blue curves respectively) collected on
2011~June 18 (cycle 84+1.466).  A clear Zeeman signature (with a full amplitude of 0.5\%
and featuring a double sign switch across the line) is detected in the LSD
Stokes $V$ profile, in conjunction with the unpolarised line profile.
The mean polarization profile is expanded by a factor of 20 and shifted upwards
by 1.1 for display purposes.  }
\label{fig:lsd}
\end{figure}

Least-Squares Deconvolution \citep[LSD,][]{Donati97b} was applied to all
observations.   The line list we employed for LSD is computed from an {\sc
Atlas9} LTE model atmosphere \citep{Kurucz93} and corresponds to a K6IV
spectral type ($\teff=4,250$~K and  $\logg=4.0$) appropriate for GQ~Lup.
As usual, only moderate to strong atomic spectral lines are included
in this list \citep[see, e.g.,][for more details]{Donati10b}.  
Altogether, about 8,900 spectral features (with about 40\% from \fei) are used
in this process.
Expressed in units of the unpolarized continuum level $I_{\rm c}$, the average
noise levels of the resulting Stokes $V$ LSD signatures range from 1.7 to
8.8$\times10^{-4}$ per 1.8~\kms\ velocity bin (median value 2.3$\times10^{-4}$).
Zeeman signatures are detected at all times in LSD profiles and in most accretion
proxies (see Sec.~\ref{sec:var});  an example LSD photospheric Zeeman signature
(collected during the 2011 run) is shown in Fig.~\ref{fig:lsd} as an
illustration.

\section{GQ~Lup}
\label{sec:gq}

Some of the important characteristics of GQ~Lup are not presently known with good 
accuracy;  in particular, the photospheric temperature \teff, usually taken as 4060~K and 
derived from the spectral type \citep[i.e., K7eV][]{Neuhauser05}, is only a crude 
approximation (not accurate to better than 250~K).  Given that the estimated mass of 
cTTSs like GQ~Lup (usually in the approximately isothermal Hayashi phase of their contraction, 
where they follow roughly vertical tracks downwards in the Hertzsprung-Russel diagram, towards 
the main sequence) mostly depends on the photospheric temperature, deriving as accurate an 
estimate of \teff\ as possible is of obvious concern for our study, especially when it comes to 
investigating how the magnetic topology relates to the internal structure of the protostar 
(see Sec.~\ref{sec:dis}).  

Towards this aim, we developed an automatic spectral classification 
tool similar to previously published ones \citep[e.g.,][]{Valenti05} and using multiple 
spectral windows in the wavelength ranges 515--520~nm and 600-620~nm.  As a first step, 
we adjust the parameters of most atomic lines in the domain to match synthetic spectra 
with those of a handful of standard stars (including the Sun) with \teff, logarithmic gravity 
(\logg, with $g$ in cgs units) and logarithmic metallicity (with respect to that of the Sun) 
in the range 4000 to 6500~K, 3.5 to 4.5 and --0.5 to 0.5;  
in a second step, we build a library of synthetic spectra for all values of \teff, \logg\ 
and logarithmic metallicity in the same range, with respective steps of 50~K, 0.1 and 0.1;  
the final step consists in fitting the observed spectrum to be characterised with all synthetic 
spectra within the grid, and deriving the stellar parameters that minimise \chisq\ 
and the corresponding error bars (from the curvature of the 3D \chisq\ landscape at the 
derived minimum);  estimates of the radial velocity \vrad\ and of the line-of-sight projected 
rotation velocity \vsini\ are also derived as a by-product, assuming a given microturbulent 
velocity (set to 1~\kms) and a tabulated macroturbulent velocity depending on \teff\ and 
\logg\ \citep[following][]{Valenti05}.  Though still in a preliminary stage, our automatic 
spectral classification tools (called MagIcS) is found to behave well on standard stars, 
yielding stellar parameters in agreement with published values within better than 50~K, 0.1 
and 0.1 for \teff, \logg\ and logarithmic metallicity.  A more detailed 
description of this tool will be presented in a dedicated paper.  

This tool was slightly modified in the specific case of cTTSs, for which optical veiling 
(i.e., the apparent weakening of the photospheric spectrum, presumably caused by accretion) 
comes as an additional parameter and makes it fairly difficult to estimate metallicity with 
reasonable accuracy at the same time.  For this purpose, we decided to carry out the fit 
assuming solar metallicity (for the grid of synthetic spectra);  we otherwise proceed in 
exactly the same way for deriving \teff\ and \logg\ (with their error bars, along with 
\vsini, \vrad\ and veiling).  When tested on standard cTTSs like BP~Tau, 
AA~Tau or V2129~Oph, MagIcS is found again to provide \teff\ and \logg\ 
compatible with published values within better than 50~K and 0.2;  applying it to 
GQ~Lup, we obtain $\teff=4,300$~K and $\logg=3.7$.  In particular, the photospheric temperature we derive is 
significantly larger than the estimate quoted in most studies (4060~K);  it is 
closer to the estimate derived from multi-colour photometry by \citet{McElwain07}, equal 
to $\simeq$4200~K.  We believe that our new measurement, based on high-\sn\ 
high-resolution spectra (see Sec.~\ref{sec:obs}) and obtained through a direct comparison 
(calibrated on standard stars) of observed and synthetic spectral lines, is more accurate 
than the two quoted published values (estimated either from the spectral type or from 
spectrophotometry only) and hence use it in the following.  

\begin{figure}
\includegraphics[scale=0.35,angle=-90]{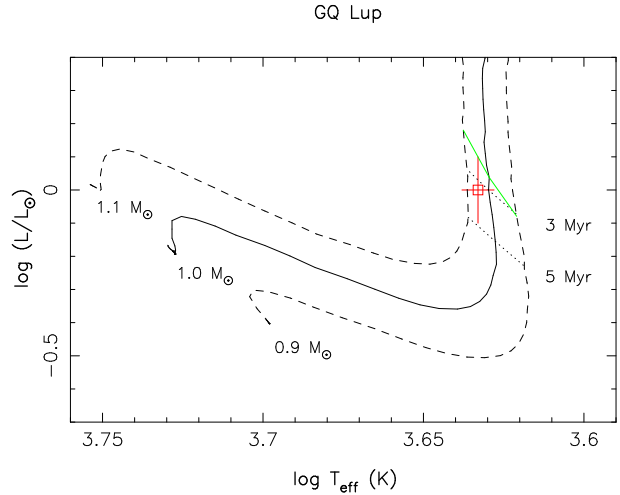}
\caption[]{Observed (open square and error bars) location of GQ~Lup in the HR diagram.
The PMS evolutionary tracks and corresponding isochrones \citep{Siess00} assume solar
metallicity and include convective overshooting.  The green line depicts where models 
predict cTTSs start developing their radiative core as they contract towards the main 
sequence.  }
\label{fig:hrd}
\end{figure}

The revised mass of GQ~Lup that we derive \citep[mostly from \teff\ and a comparison with 
evolutionary models of][see Fig.~\ref{fig:hrd}]{Siess00} is $1.05\pm0.07$~\msun\ (the error 
bar reflecting mainly the uncertainty on \teff).  This 
is significantly larger than most estimates published so far \citep[usually in the range 
0.7--0.8~\msun, e.g.,][essentially reflecting the difference in \teff]{Neuhauser05,Seperuelo08} 
but more in line with the findings of \citet[][also favouring a larger \teff]{McElwain07}.  
In addition to small ($<$0.5~mag) photometric fluctuations attributed as usual to the presence 
of cool spots coming and going on the visible hemisphere of the protostar \citep[e.g.,][]{Broeg07}, 
GQ~Lup is also reported to undergo much larger dimming episodes \citep[of up to 2~mag, 
e.g.,][]{Covino92} that can hardly be caused by cool starspots but more likely by extinction 
(e.g., by circumstellar material passing onto the line of sight);  
by considering only epochs at which the photometric brightness is highest and the 
photometric variability is smallest \cite[as in, e.g.,][]{Janson06}, one can derive an 
estimate of the unspotted, unextincted, $V$ brightness of GQ~Lup,
equal to $V=11.4\pm0.2$ \citep[e.g.,][]{Broeg07}.  Using the above mentioned distance 
($150\pm10$~pc) as well as the bolometric correction corresponding to our temperature 
estimate \citep[$-0.74\pm0.05$, e.g.,][]{Bessell98}, we finally obtain that GQ~Lup has a 
bolometric magnitude of $4.78\pm0.25$ (corresponding to a logarithmic luminosity with respect to the 
Sun of $\log \lstar/\lsun=0.0\pm0.1$, see Fig.~\ref{fig:hrd}) and thus a radius of $1.7\pm0.2$~\rsun\ \citep[in 
reasonable agreement with the findings of][despite the different values of \teff\ 
used in both studies]{Seperuelo08}.  The (conservative) error bar on the radius mostly 
reflects those on the luminosity and temperature.  
By comparing to evolutionary 
models of \citet[][see Fig.~\ref{fig:hrd}]{Siess00}, we find that GQ~Lup has an age of 
2--5~Myr, in good agreement with the estimates of \citet{McElwain07} and \citet{Seperuelo08};  
evolutionary models 
also predict that, at this age, GQ~Lup should no longer be fully-convective and should host 
a radiative core extending up to $\simeq$0.25~\rstar\ in radius \citep{Siess00}.

\section{Spectroscopic variability}
\label{sec:var}

Before carrying out a full modelling of our spectropolarimetric data (see Sec.~\ref{sec:mod}), 
we first present a simple phenomenological description of how the photospheric LSD profile and 
the selected accretion proxies (i.e., \caii\ IRT, \hei\ $D_3$, \hal\ and \hbe) vary with 
time over each of our 2 observing runs.  More specifically, this first step consists 
in looking at how equivalent widths, RVs and longitudinal magnetic fields (i.e., the 
line-of-sight projected component of the vector field averaged over the visible stellar
hemisphere and weighted by brightness inhomogeneities) are modulated by rotation, in 
order to derive a rough, mostly intuitive, idea about the nature and orientation 
of the large-scale field of GQ~Lup, and about the surface distribution of cool 
photospheric spots and hot chromospheric accretion regions.  

\subsection{The rotation period}

A recent search for photometric (and RV) modulation by \citet{Broeg07} has suggested that
the rotation period of GQ~Lup is $8.45\pm0.2$~d, typical of that of cTTSs in the mass range
0.5--1.2~\msun\ and with ages of $<$5~Myr.  Clear modulation at this period is found in
their data (at 2 different epochs), as well as in several (though not all) older photometric
light curves published in the literature over the last 2 decades;  this modulation is
potentially also present in the RV estimates derived with cross-correlation from a series
of high-resolution spectra.  Modulation of the amount of veiling in the spectrum of GQ~Lup
is also reported by \citet{Seperuelo08} from spectrophotometric observations covering a
timespan of 17~d;  the period they report is however significantly longer, equal to
10.7~d, compatible with the one they determine from archival B band photometry
($10.43\pm0.12$~d).  This apparent discrepancy may be caused by intrinsic variability 
(rendering rotation period of cTTSs difficult to estimate precisely) or reflect something real, 
such as differential rotation, at the surface of GQ~Lup.  
As detailed in the following subsections, our own data (and in particular the 2009 data set) 
also show clear rotational modulation,
both in the circularly polarized and the unpolarized profiles of photospheric lines and
accretion proxies) at an average period of $8.4\pm0.3$~d.  
We therefore selected this period as the main rotation period of GQ~Lup, and used it to 
phase all our data (see Sec.~\ref{sec:obs}).  

Given the \vsini\ we estimate (equal to
$5\pm1$~\kms, see Sec.~\ref{sec:mod}), we derive (from the radius and rotation period) that
the rotation axis of GQ~Lup is inclined at $\simeq$30\degr\ to the line of sight, and hence
that GQ~Lup is seen mostly pole-on (rather than equator on), in agreement with the findings
of \citet{Broeg07} and with a recent estimate of the viewing angle of the accretion 
disc from interferometric data (Anthonioz, private communication).  

\begin{figure*}
\center{\includegraphics[scale=0.80,angle=-90]{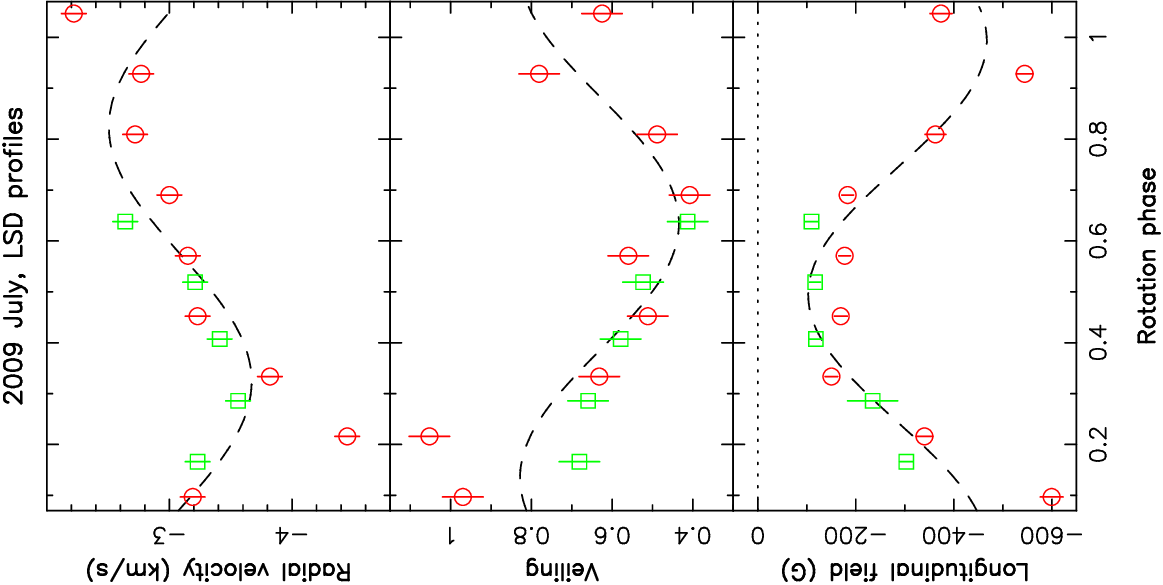}\hspace{4mm}
\includegraphics[scale=0.80,angle=-90]{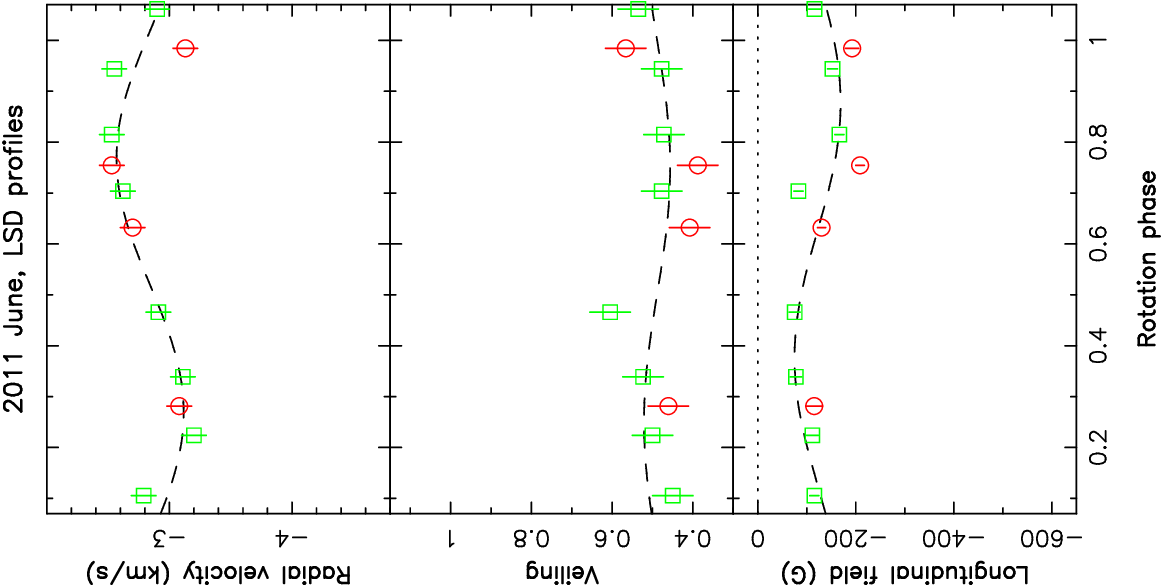}}
\caption[]{Rotational modulation of the RV (top row), veiling (second row)
and longitudinal field (bottom row) derived from the LSD photospheric profiles
of GQ~Lup in 2009~July (left panels) and in 2011~June (right panels).
Data collected during the first and second rotational cycles of each run (i.e., 
corresponding to phases 0.07--1.07 and 1.07--2.07 respectively) are shown with 
red circles and green squares.  Formal $\pm$1~$\sigma$ error bars (computed from the error bars 
of the observed spectra) are shown for longitudinal fields, while conservative error 
bars of $\pm$0.1~\kms\ and $\pm$0.05 were assumed for RVs and veiling respectively.  
Fits with sine/cosine waves are included (and shown as dashed lines) to outline
(whenever significant) the amount of variability attributable to rotational 
modulation.   }
\label{fig:var1}
\end{figure*}

\begin{figure*}
\center{\includegraphics[scale=0.80,angle=-90]{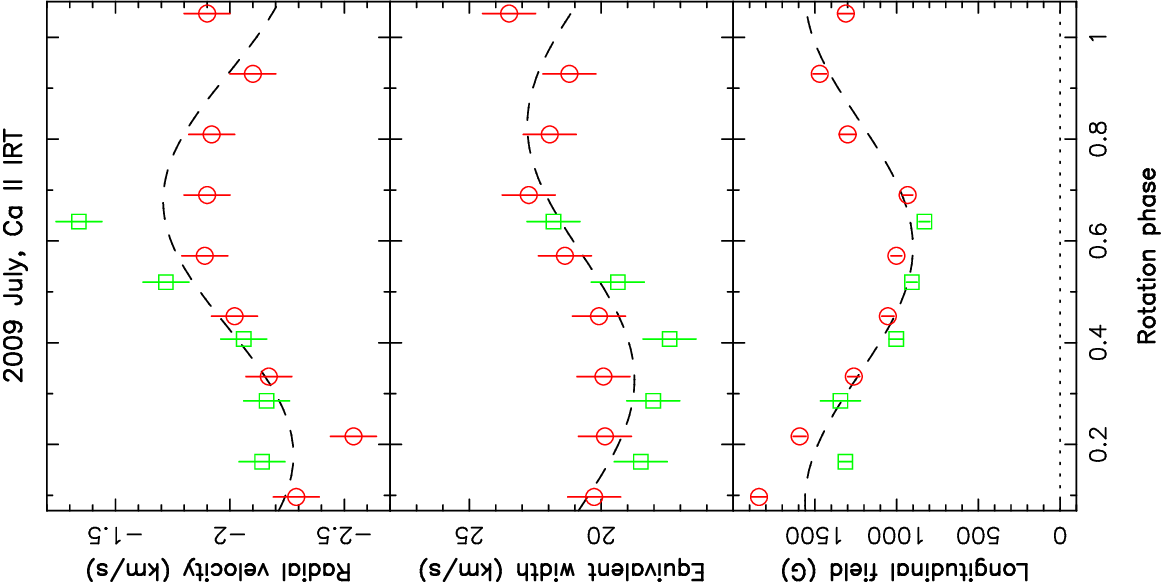}\hspace{4mm}
\includegraphics[scale=0.80,angle=-90]{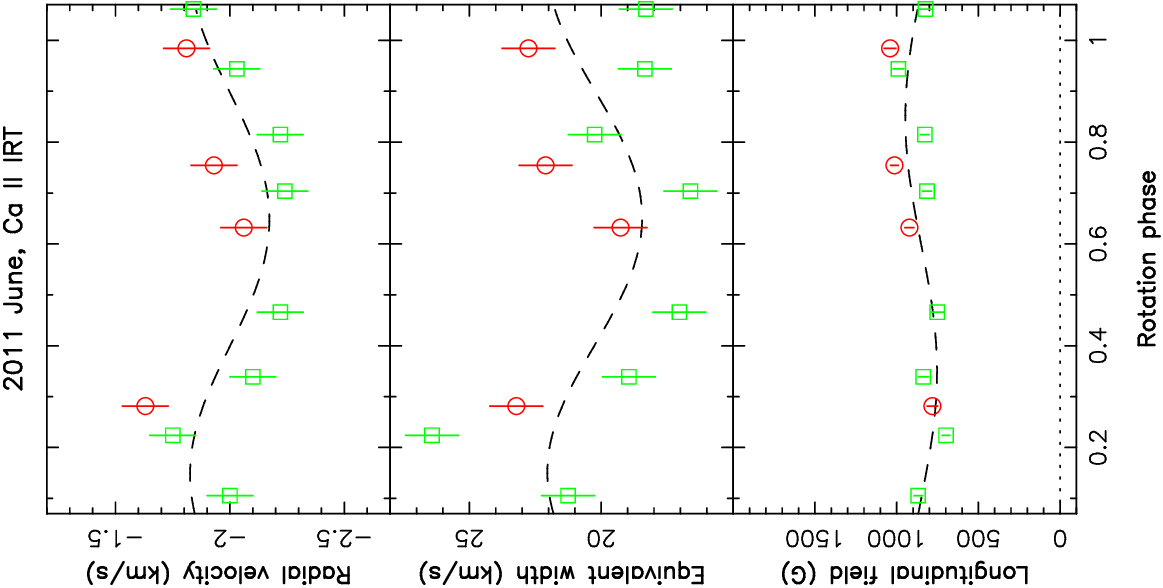}}
\caption[]{RVs (top row), equivalent widths (second row) and longitudinal fields (bottom row) 
derived from the \caii\ IRT LSD profiles of GQ~Lup in 2009~July (left panels) and in 2011~June 
(right panels), with symbol/color coding as in Fig.~\ref{fig:var1}. 
Conservative error bars of $\pm$0.1 and $\pm$1~\kms\ were assumed for the RVs and equivalent widths 
of the emission profile.  }
\label{fig:var2}
\end{figure*}

\subsection{LSD photospheric profiles}

We first examine the temporal variability of unpolarized and circularly polarized LSD profiles 
of GQ~Lup, summarised graphically in Fig.~\ref{fig:var1} for both observing epochs.  
In 2009~July (left column of Fig.~\ref{fig:var1}), RVs of Stokes $I$ LSD profiles show clear 
and smooth rotational modulation about a mean RV of $\vrad=-3.2\pm0.1$~\kms\ and 
with a full amplitude of $\simeq$1.2~\kms\ that repeats quite well between the 2 observed 
cycles, except for 2 stray points at rotation cycles 0.216 and 1.047 that are typically off 
by $\simeq$1~\kms\ from the average trend.  
In 2011~June (right column of Fig.~\ref{fig:var1}), Stokes $I$ LSD profiles are on average 
slightly (but significantly) red-shifted with respect to those of 2009~July 
($\vrad=-2.8\pm0.1$~\kms) and exhibit a smaller level of 
rotational modulation (full RV amplitude of $\simeq$0.5~\kms), but are otherwise very 
similar and repeat in particular rather well between the 2 observed cycles.  
It is reasonable to assume that the rotational modulation we detect is mostly caused by the 
presence of cool spots at the surface of GQ~Lup, as for all other cTTSs already studied in 
our sample \citep[e.g.,][]{Donati11}.  The small amplitude of the RV rotational 
modulation (relative to the line-of-sight projected rotational velocity, $\vsini\simeq5$~\kms) 
suggests that the main cool spot causing the RV changes is located at high-latitudes (as for 
most other cTTSs), and even more so in 2011~June than in 2009~July;  moreover, the parent 
spot is presumably centred around phase 0.0--0.1 (i.e., at mid-phase between RV maximum and 
minimum).  
The period on which RVs fluctuate is found to be equal to $8.8\pm0.6$~d and $9.2\pm0.6$~d 
in 2009~July and 2011~June respectively, i.e., marginally longer than (though still compatible with) 
the assumed rotation period on which all data were phased (equal to 8.4~d, see Eq.~\ref{eq:eph}).  
The clear change in the 
RV curve between the 2 epochs demonstrates that the spot configuration significantly evolved 
between the 2 epochs, as opposed to what is seen on TW~Hya \citep[where the high-latitude spot 
configuration remained more or less stable for the last 3~yrs,][]{Donati11b}.  
The origin of the small shift in \vrad\ (about $-0.4$~\kms\ between 2009 and 2011) is unclear, 
and unlikely attributable to the distant brown dwarf companion previously reported GQ~Lup~B 
(too distant to generate such a change in as little as 2~yr).  

The spectrum of GQ~Lup is also significantly veiled at both our observing epochs, with average 
veiling (at about 640~nm, the average wavelength of LSD profiles) varying from 0.4 to 1  
(see middle panels of Fig.~\ref{fig:var1}).  
Veiling variations are weak in 2011~June ($\simeq$10\% peak to peak about an average veiling of 
$\simeq$0.5) but significantly stronger in 2009~July where veiling is clearly modulated by 
rotation (by $\simeq$40\% peak to peak, about a mean of $\simeq$0.65 and with a period of 
$8.2\pm0.6$~d), suggesting that GQ~Lup was in a more active state of accretion in 2009~July than 
in 2011~June.  
In 2009~July, veiling is found to peak around phase 0.1, i.e., when the cool polar spot 
detected at the surface of GQ~Lup (through RV changes) comes closest to the observer.  
This suggests that the hot chromospheric spot presumably causing the veiling variations 
roughly overlaps the cool photospheric spot, e.g., as in most other cTTSs we studied to date.  

Zeeman signatures of GQ~Lup are detected at all times, 
showing in most cases a canonical shape (i.e., antisymmetric with respect to the line centre) 
with a peak-to-peak amplitude ranging from 0.3 to 1.1\%, except at a few rotational cycles 
(e.g., cycle 84+1.466, see Fig.~\ref{fig:lsd}) where the Zeeman signature exhibits a more 
complex shape (e.g., featuring a double sign switch across the line profile).  
The corresponding longitudinal field is found to be negative at all times, ranging from 
--75 to --600~G (with typical error bars of 10--20~G).  
As for RVs, longitudinal fields show clear rotational modulation (see bottom panels of 
Fig.~\ref{fig:var1}), especially in 2009~July 
where the peak-to-peak fluctuation reaches almost 400~G with a well defined period of 
$8.4\pm0.3$~d (the period that we selected for phasing all our data, see Eq.~\ref{eq:eph}).  
The longitudinal field of GQ~Lup is found to reach maximum strength around phase 0.9-1.0, 
i.e.\ slightly leading the high-latitude cool spot tracked through RV changes (best 
viewed at phase $\simeq$0.1).  The drastic change in the amplitude of the 
longitudinal field curve between 2009 and 2011 also indicates that the large-scale field 
of GQ~Lup has significantly evolved between our two observing epochs.  
In addition to rotational modulation (dominant in 2009), intrinsic variability is also 
clearly detected at both epochs and likely reflects local, rapid changes at the surface of 
GQ~Lup (e.g., in the photospheric brightness distribution) possibly resulting 
from unsteady accretion.   

\subsection{\caii\ IRT emission}

We use core emission in the \caii\ IRT lines as our main proxy of surface accretion;  
despite the inconvenience of including a dominant contribution from the non-accreting 
chromosphere, this proxy demonstrates a number of clear advantages over the more 
conventional \hei\ $D_3$ accretion proxy.  In particular, \caii\ IRT lines are formed 
closer to the stellar surface and in a more static atmosphere, making them easier (and 
less-model dependent) to interpret in terms of large-scale magnetic topologies, with 
no need to model the velocity flow in the line formation region;  they 
also provide higher quality data which usually overcompensates the signal dilution from 
the non-accreting chromospheric regions.  
We start by building up a \caii\ IRT emission profile for each of our spectra;  we achieve 
this by constructing LSD-like weighted averages of the 3 IRT lines, then by subtracting 
the underlying (much wider) Lorentzian absorption profiles, with a single Lorentzian fit 
to the far line wings \citep[see][for an illustration]{Donati11b}.  As for photospheric 
LSD profiles, we finally examine how the RVs, equivalent widths and longitudinal fields of 
these emission profiles vary with time throughout our 2 observing runs;  these variations 
are graphically summarised in Fig.~\ref{fig:var2}.  

We first note that \caii\ emission strengths are roughly equal at both epochs, with 
average equivalent widths of $\simeq$21~\kms\ (or 0.060~nm, see middle panels 
of Fig.~\ref{fig:var2}).  Rotational modulation is only moderate (smaller than 25\% 
peak-to-peak) but clear in 2009~July (where the modulation period is found 
to be $8.9\pm0.7$~d, larger though still compatible with that assumed in Eq.~1), 
while intrinsic variability can be significant at times (e.g., 
2011~June).  Moreover, \caii\ emission profiles exhibit regular RV fluctuations about 
a mean of $-2.0\pm0.1$~\kms\ in 2009~July and $-1.9\pm0.1$ in 2011~June, i.e., 
red-shifted by typically $0.9-1.2$~\kms\ with respect to the photospheric spectrum 
\citep[as observed so far for moderately accreting cTTSs, see, e.g.,][and references therein]{Donati11b}.  
The full amplitude of the RV rotational modulation is small (with respect to 
the rotational line broadening), of order 0.5~\kms\ at both epochs, suggesting that 
the region of excess \caii\ emission causing this modulation is located at 
high-latitudes;  the period of the RV modulation is found to be $8.1\pm0.5$ and 
$8.3\pm0.9$~d in 2009~July and 2011~June respectively, in agreement with the rotation 
period used to phase our data.  

In 2011~June (right column of Fig.~\ref{fig:var2}), we note that \caii\ emission RV 
variations are mostly anti-correlated with those of LSD profiles, as often the case for moderately 
accreting cTTSs \citep[e.g., TW~Hya,][]{Donati11b} and naturally expected when chromospheric 
regions of \caii\ excess emission are roughly co-spatial with cool photospheric spots.  
Whereas this modulation suggests that the parent accretion region is centred at phase 
$\simeq$0.9 (i.e., midway between RV minimum and maximum), \caii\ excess emission is 
found to reach maximum around phase 0.15 (though with a high-level of intrinsic 
dispersion, see Fig.~\ref{fig:var2}, right column, second graph);  this is roughly 
compatible with the position of the photospheric cool spot as derived from LSD 
photospheric profiles (around phase 0.0--0.1), phase delays as large as 
0.1 rotation cycle being much easier to produce (and therefore not necessarily very 
significant) at high latitudes (where they correspond to shorter physical distances) 
than at low latitudes.  

The situation is less clear in 2009~July.  Whereas \caii\ emission strength peaks 
around phase 0.85, the corresponding RVs suggest that the accretion spot is 
located around phase 0.45 (i.e.\ half way between phases of RV minimum and maximum);  
in particular, RVs of \caii\ emission are largely correlated (in fact, shifted by 
$\simeq$0.15 rotation cycle) with those of LSD profiles, rather than being 
anti-correlated as in most other cases (e.g., on GQ~Lup in 2011~June).  
At this point, we have no clear idea as to why RVs of \caii\ emission behave in this 
unusual way at this specific epoch;  this is the first such example that we have 
encountered among the $\simeq$5 cTTSs studied in detail so far.  
While we need to gather more similar observations to progress on this issue, 
we recall for now that the observed RV variations, although weird, are 
low in amplitude (0.5~\kms\ peak to peak), thus confirming at the very least that 
the corresponding feature is located close to the pole.  

As for LSD profiles, large Zeeman signatures are detected at all times  
within the emission component of \caii\ IRT lines, with peak-to-peak 
relative amplitudes of up to 25\% in 2009~July.  
The corresponding longitudinal fields, ranging from 0.8 to 1.9~kG with median 
error bars of about 35~G (see lower panels of Fig.~\ref{fig:var2}), are the 
strongest among all cTTSs investigated up to now.  Rotational 
modulation is clear and well sampled in 2009~July (with an average full amplitude of 
$\simeq$700~G, and a modulation period of $8.0\pm0.4$~d) but much weaker in 
2011~June (full amplitude lower than 200~G) where the longitudinal field hardly 
exceeds 1~kG;  longitudinal field maximum 
is reached at phase 0.1 in 2009~July and $\simeq$0.9 in 2011~June, confirming 
that the cool photospheric spot and the chromospheric accretion region (also 
located at about the same phase given the above mentioned proxies) are roughly 
cospatial with the main magnetic poles as in most cTTSs analysed up to now.  
Although present at both epochs, intrinsic variability of longitudinal fields 
is only moderate on GQ~Lup.  

As for several other cTTSs, we note the striking difference in magnetic polarity 
between the (constantly positive) longitudinal fields measured from the \caii\ emission 
lines and those (always negative) derived from the LSD profiles, which again 
demonstrates that LSD profiles and IRT lines probe different areas (of 
different magnetic polarities) over the stellar surface;  following previous papers
\citep[e.g.,][]{Donati11}, we suggest that this is once more evidence that the 
field of GQ~Lup is mostly octupolar, with \caii\ IRT lines (and other accretion proxies) 
probing mostly the high-latitude (positive) magnetic pole and photospheric LSD 
profiles reflecting essentially the low-latitude belt of (negative) radial field.  
The strong weakening from 2009 to 2011 of both the average longitudinal field and 
of the amount of rotational modulation (also detected on LSD photospheric Zeeman 
signatures) further confirms that the large-scale field has evolved 
between our 2 observing epochs.  

\begin{figure*}
\center{
\includegraphics[scale=0.65,angle=-90]{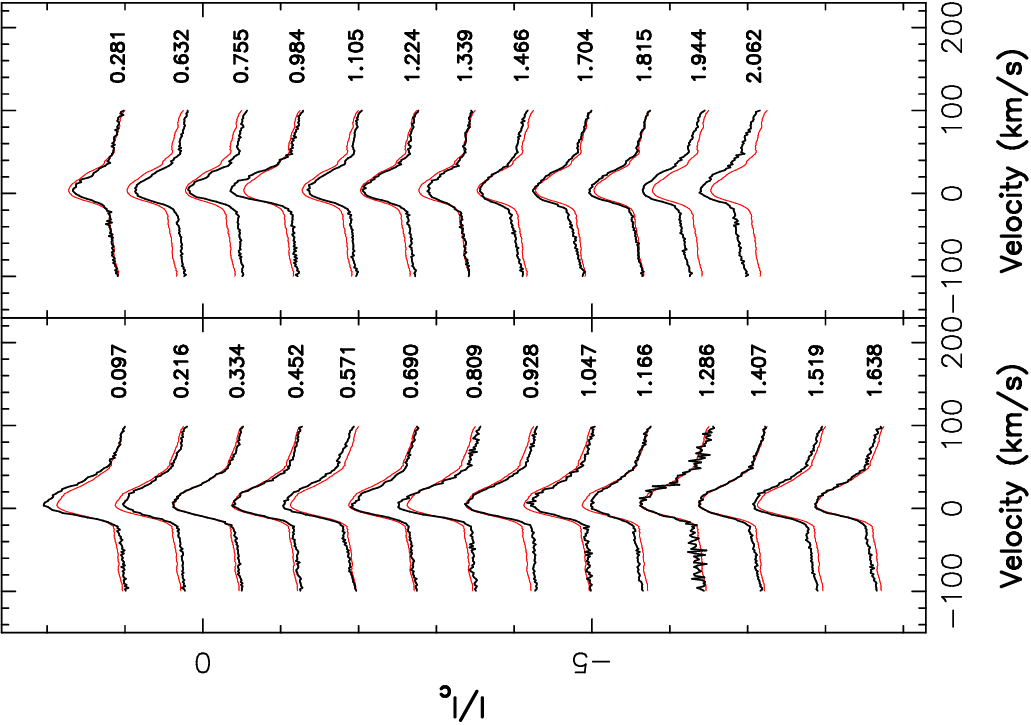}\hspace{5mm}
\includegraphics[scale=0.65,angle=-90]{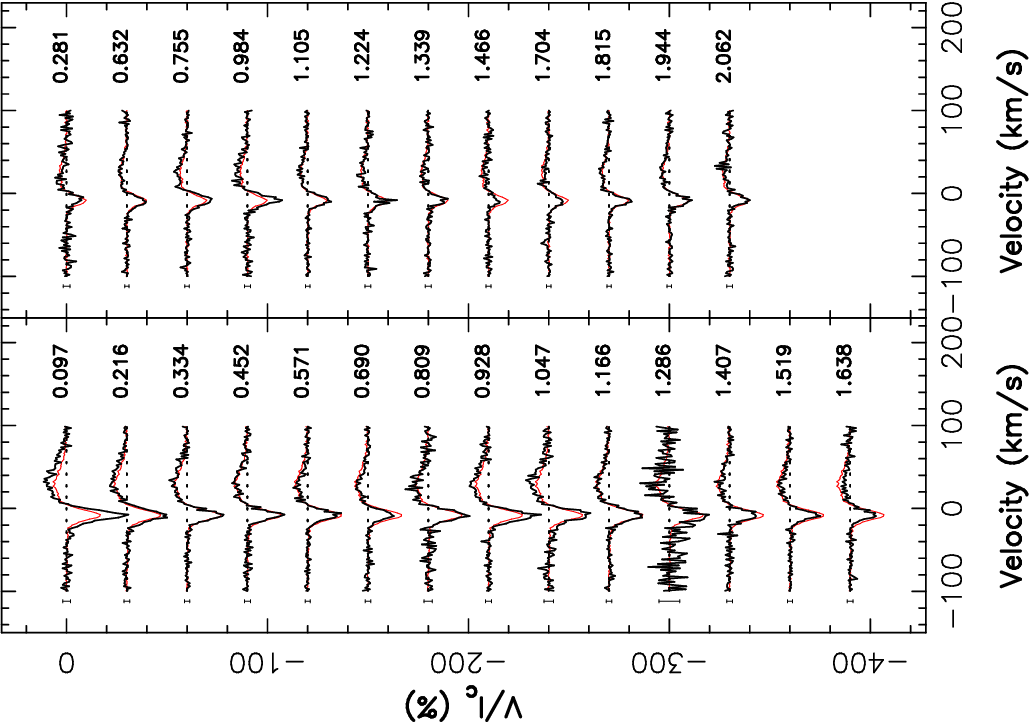}}
\caption[]{Variations of the unpolarized (Stokes $I$, left panel) and circularly-polarized
(Stokes $V$, right panel) profiles of the \hei\ $D_3$ emission of GQ~Lup in 2009~July (left
columns of both panels) 2011~June (right columns).  
Large Zeeman signatures (with full amplitudes of up to 40\% in 2009~July) 
are clearly detected at all epochs and exhibit shapes that strongly depart from the usual 
antisymmetric pattern, with a strong / narrow blue (negative) lobe and a weak / wide red (positive) lobe.  
To emphasize variability, the average profile over each run is shown in red.
Rotation cycles (as listed in Table~1) and 3$\sigma$ error bars (for Stokes $V$ profiles only)
are shown next to each profile.  }
\label{fig:he}
\end{figure*}

\subsection{\hei\ $D_3$ emission}

\begin{figure*}
\center{\includegraphics[scale=0.80,angle=-90]{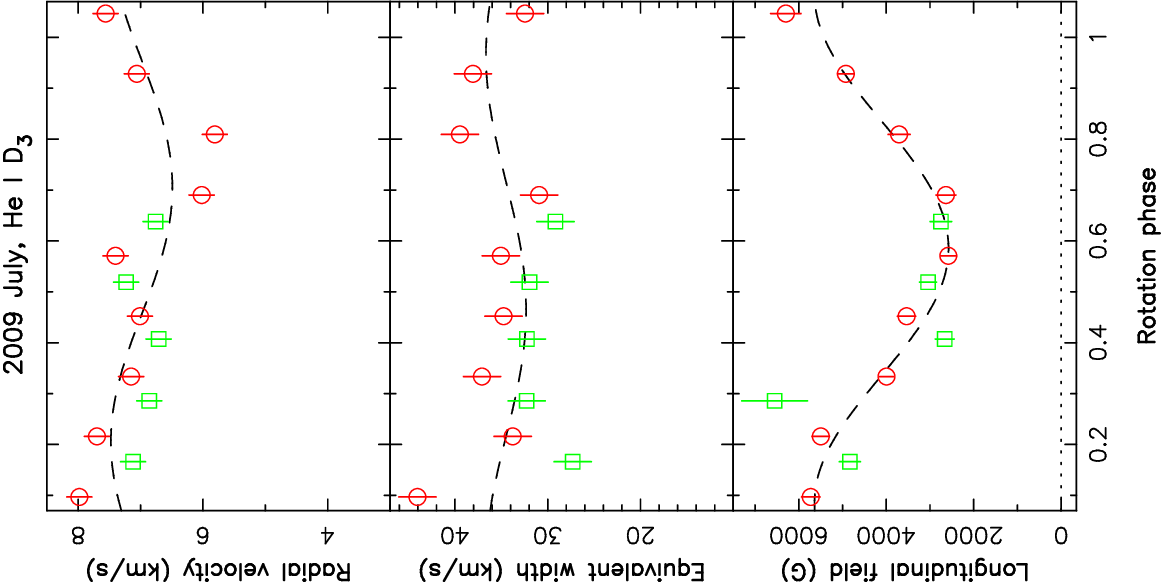}\hspace{4mm}
\includegraphics[scale=0.80,angle=-90]{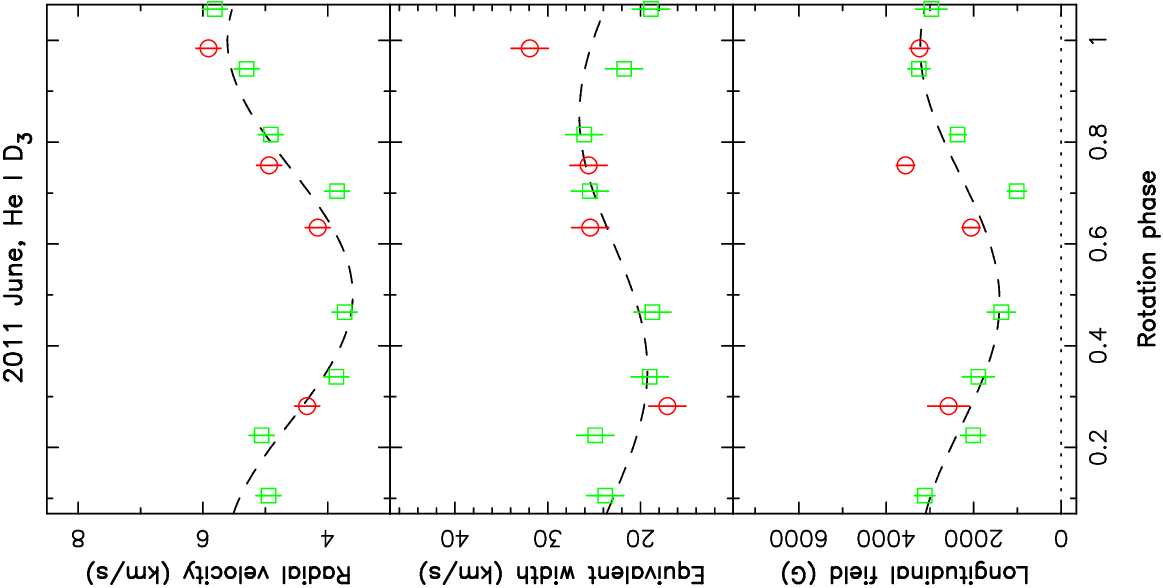}}
\caption[]{Same as Fig.~\ref{fig:var2} for the narrow component of the \hei\ $D_3$ profiles of GQ~Lup. 
Conservative error bars of $\pm$0.2~\kms\ and $\pm$2~\kms\ were assumed on the RVs and equivalent widths 
of the emission profile.  }
\label{fig:var3}
\end{figure*}

During our 2 observing runs, the \hei\ $D_3$ line of GQ~Lup (see Fig.~\ref{fig:he}) showed up 
most of the time as a narrow emission profile (with a typical full width at half maximum of 
$\simeq$40~~\kms), sometimes on top of a much broader emission profile (e.g., on rotational 
cycles 1.944 and 2.062) as for TW~Hya \citep{Donati11b};  here we only consider the narrow 
emission, presumably probing the postshock zone of the accretion region where the plasma is 
experiencing a strongly decelerating fall towards the stellar surface.  We find that this 
narrow emission is centred on $\simeq$7~\kms\ in 2009~July and on $\simeq$4.5~\kms, i.e., 
shifted with respect to the photospheric lines by 7-10~\kms\ depending on the epoch (making 
GQ~Lup similar to other cTTSs in this respect).  

The RV and equivalent width variations of the \hei\ $D_3$ 
emission profile are both compatible with (and thus attributable to) the presence of a hot 
accretion spot best visible at phases 1.0 and 0.8 for epochs 2009~July and 2011~June 
respectively (i.e., at maximum equivalent width, and midway between RV minimum and maximum, 
see Fig.~\ref{fig:var3}, top and middle panels);  this is in rough agreement with 
previous conclusions from both LSD photospheric and \caii\ IRT emission profiles.  
Since the amplitude of the RV variations are small ($<2$~\kms\ at both epochs), it again 
suggests that this accretion spot is located close to the pole.  The modulation periods of 
emission strengths are found to be marginally shorter (i.e., $7.3\pm1.2$ and $7.6\pm0.5$~d 
in 2009~July and 2011~June respectively) than the assumed rotation period, whereas those 
from RVs are found to be somewhat longer (i.e., $8.8\pm1.3$ and $9.0\pm0.3$~d).  

We note that the \hei\ emission strength decreased significantly between 2009~July and 
2011~June (from 35 to 25~\kms, or equivalently from 0.070 to 0.050~nm, see Fig.~\ref{fig:var3}), 
in agreement with the veiling of photospheric lines (also stronger in 2009~June, 
see Fig.~\ref{fig:var1}).  However, no such decrease is seen in the emission core of 
the \caii\ line, whose strength is roughly the same at both epochs.  
Since only a small fraction ($\simeq$25\%, see Sec.~\ref{sec:mod}) of the \caii\ 
emission flux is attributable to accretion, we suspect that the change observed in \hei\ 
emission and veiling is likely occurring as well in the flux of the \caii\ emission core, 
but at a low enough level ($\simeq$10\% of the overall \caii\ emission flux, if the fraction 
attributable to accretion varied in the same proportion as the \hei\ emission flux) to 
remain hidden by, e.g., epoch-to-epoch fluctuations in the (dominant) remaining fraction 
of the emission flux due to chromospheric activity.  

Large Zeeman signatures (with full amplitudes of up to $\simeq$40\%) are observed in 
conjunction with the narrow emission component (see Fig.~\ref{fig:he}, right panel), probing 
longitudinal fields ranging from 1 to more than 6~kG (see Fig.~\ref{fig:var3}, lower panels) 
making GQ~Lup the cTTSs with strongest magnetic fields known to date.  
As observed so far in cTTSs, the shape of \hei\ Zeeman signatures (whenever detected) strongly 
departs from the usual 
antisymmetric pattern (with respect to the line centre, see Fig.~\ref{fig:he}) confirming 
that the line forms in a region where the accreted plasma is rapidly decelerating.  

Rotational 
modulation and polarity of longitudinal fields are very similar to those of the \caii\ line, 
demonstrating that both lines are reliable probes of magnetic fields in the accretion 
region.  Intrinsic variability in longitudinal fields (on timescales of days) is larger than 
formal error bars but nevertheless moderate at both epochs (and in particular lower than the 
amount of rotational modulation), as already noticed on photospheric LSD profiles and \caii\ 
emission lines.  
The clear weakening of the longitudinal field between 2009~July and 2011~June 
further demonstrates that the large-scale magnetic topology is subject to very significant 
intrinsic variability on timescales as short as 2~yr.  
The longitudinal field modulation periods are found to be $7.9\pm0.3$ and 
$10.1\pm0.5$~d in 2009~July and 2011~June respectively.  We do not think that differential 
rotation is likely to explain such variations, since \hei\ emission is obviously coming 
from near the pole at both epochs, i.e., from regions too close in latitude to generate 
the observed change in modulation period;  we rather suspect intrinsic variability 
(coupled to a reduced modulation amplitude and a limited coverage) to be causing the 
unusually large (i.e., 
3.4$\sigma$) discrepancy between the estimated modulation period in 2011 and the average 
period of 8.4~d (taken as the rotation period).  

\begin{figure*}
\center{
\includegraphics[scale=0.65,angle=-90]{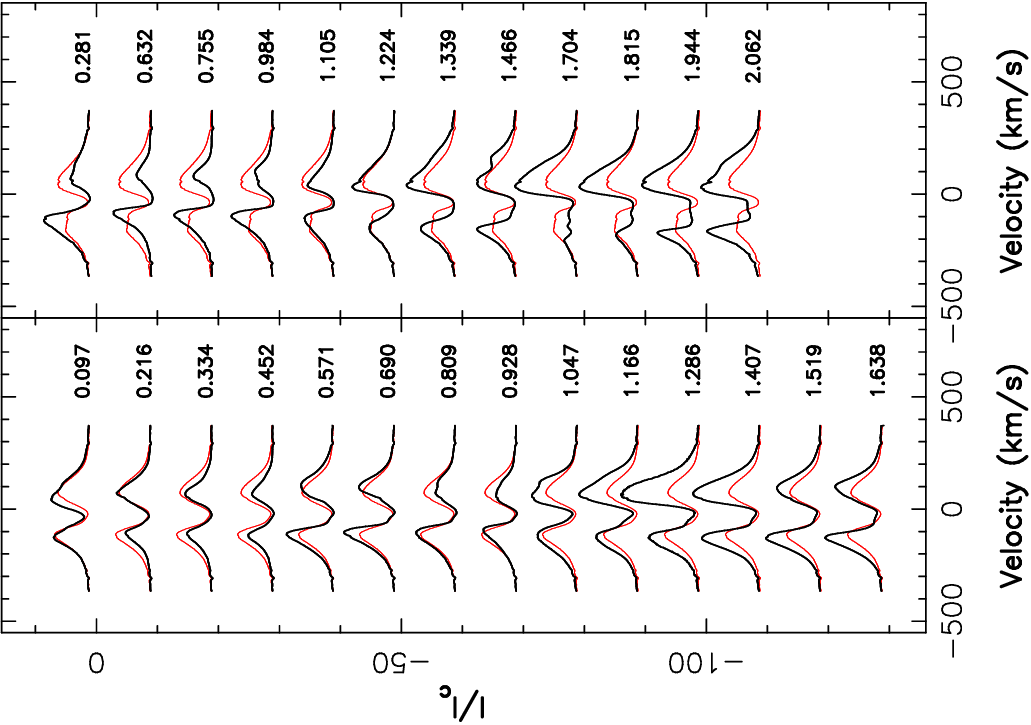}\hspace{5mm}
\includegraphics[scale=0.65,angle=-90]{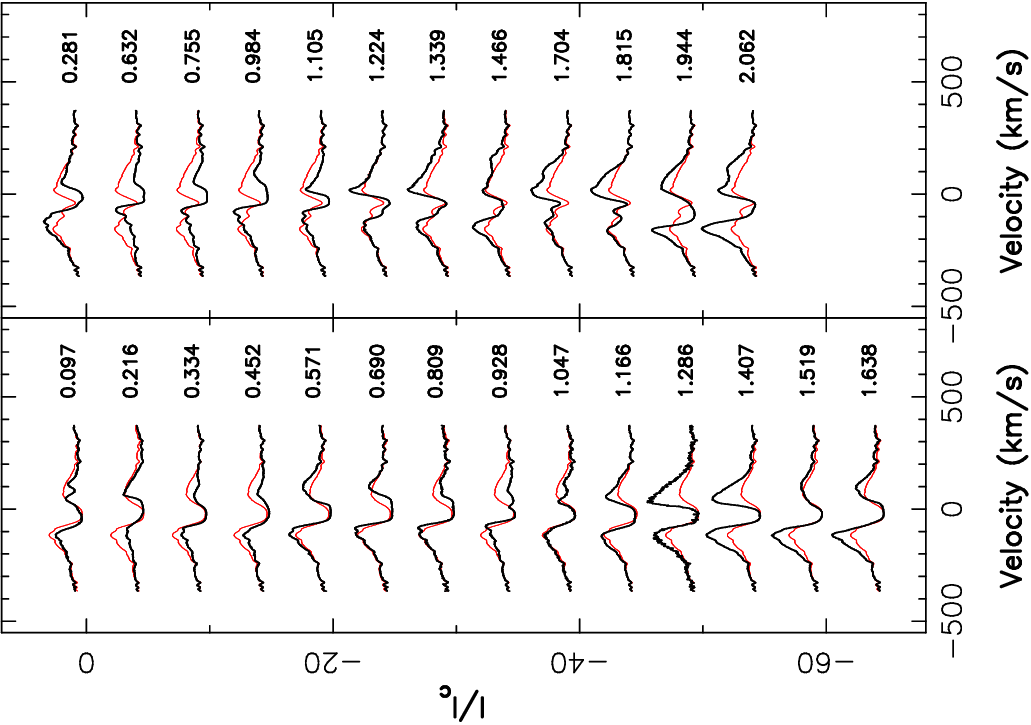}}
\caption[]{Variations of the \hal\ (left) and \hbe\ (right) lines in the spectrum of
GQ~Lup, in 2009~July (left column of both panels) and 2011~June (right column).
To emphasize variability, the average profile over each run is shown in red.
Rotation cycles (as listed in Table~1) are mentioned next to each profile.  }
\label{fig:bal}
\end{figure*}

\subsection{Balmer emission}

Unsurprisingly, Balmer lines of GQ~Lup are in emission (see Fig.~\ref{fig:bal}), with 
\hal\ and \hbe\ both showing conspicuous double-peak profiles reminiscent of those of AA~Tau 
\citep[e.g.,][]{Bouvier07b, Donati10b};  these profiles are in particular more complex 
than the inverse P~Cygni profiles previously reported for GQ~Lup from low-resolution 
spectrophotometric observations \citep[][]{Batalha01}, in agreement with more recent 
(also low-resolution spectrophotometric) observations \citep[][]{Seperuelo08}.  
Their respective average equivalent widths are equal to 
1,500 \& 250~\kms\ (3.3 \& 0.40~nm) in 2009~July and to 1,600~\kms\ \& 400~\kms\ 
(3.5 \& 0.65~nm) in 2011~June;  as a result of the central (blue-shifted) absorption, 
likely due to winds rather than to accretion \citep[e.g.,][]{Seperuelo08}, we suspect 
that these fluxes are lower limits only, rather than true fluxes.  

Both lines exhibit strong fluctuations with time, both in their equivalent widths and 
shape (see Fig.~\ref{fig:bal}).  However, these variations do not really correlate well 
with rotation phase and are thus unlikely attributable to rotational modulation;  
for instance, the clear strengthening of the red emission peak observed at rotational 
cycles 1.166--1.407 (in 2009~July) and 1.704--2.0662 (in 2011~June) are not observed 
one cycle before at both epochs.  Note that Balmer line emission is stronger in 2011 
than in 2009, and thus does not scale up with the amount of accretion, presumably 
strongest in 2009 (according to veiling and \hei\ emission fluxes).  

The only clear temporal behaviour we can report on 
Balmer lines (from analyses of autocorrelation matrices) is that the blue wing of the 
central (blue-shifted) absorption strongly correlates (in both \hal\ and \hbe) with 
the overall profile emission strength in 2011~June, this blue wing being strongly 
blue-shifted (with respect to the average profile) when the emission strength is stronger 
(e.g., at rotational cycles 84+1.704--2.0662).  This may suggest that winds from GQ~Lup 
(presumably causing the central absorption) are also getting faster when they get stronger.  

No true absorption is found to occur in the red wing of Balmer lines, and \hbe\ in 
particular, like those found for AA~Tau \citep{Bouvier07b, Donati10b} and for V2129~Oph 
\citep{Donati11, Alencar12}.  As a consequence, no unambiguous constraint can be derived 
from Balmer lines on the phases at which accretion funnels cross the line of sight (as 
for both AA~Tau and V2129~Oph);  this is not really surprising though, given the fact 
that GQ~Lup is viewed mostly pole-on rather than equator-on (see Sec.~\ref{sec:gq}), 
this viewing configuration being obviously less favourable for detecting the AA~Tau-like  
red-shifted absorption events in Balmer lines thought to be probing episodic crossings of 
accretion funnels.  

\subsection{Mass-accretion rate}

From the average equivalents widths of the \caii\ IRT, \hei\ and \hbe\ lines of GQ~Lup, 
and approximating the stellar continuum by a Planck function at a temperature of 4,300~K (see 
Sec.~\ref{sec:gq}), we can derive logarithmic line fluxes (with respect to the luminosity of the Sun \lsun),
respectively equal to $-4.8$, $-4.8$ and $-3.9$.  
This implies logarithmic accretion luminosities 
(with respect to \lsun) respectively equal to $-2.0$, $-1.6$ and $-2.0$ \citep[using empirical 
correlations from][]{Fang09}.  We thus conclude that the
average logarithmic mass accretion rate of GQ~Lup (in \mspy) is equal to $-9.0\pm0.3$, the 
difference in accretion rates between our 2 observing epochs ($\simeq$0.2~dex) being smaller than 
the quoted error bar.  

Mass accretion rates can in principle also be estimated (though less accurately) through the full
width of H$\alpha$ at 10\% height \citep[e.g.,][]{Natta04, Cieza10}.  In the case of GQ~Lup,
\hal\ shows a double-peak profile with a full width at 10\% height of $500\pm20$~\kms\ on average;  
the logarithmic mass accretion rate estimated from the above mentioned correlation is $-8.0\pm0.6$ 
(in \mspy), only marginally consistent with the above estimate.  The origin of this apparent 
discrepancy \citep[not seen for other similar cTTSs, e.g.,][]{Curran11} is not clear yet, 
but could relate to the unusual shape of Balmer lines of GQ~Lup (and to their central absorption 
in particular, possibly leading to an overestimate of the profiles full width at 10\% height).

\section{Magnetic modelling}
\label{sec:mod}

In this second phase of the analysis, we aim at converting our sets of photospheric LSD and \caii\ 
IRT emission profiles into maps of the large-scale magnetic topology, as well as distributions of 
surface cool spots and of chromospheric accretion regions, at the surface of GQ~Lup;  for this, 
we use automatic tools to ensure that our final conclusions are not biased by any preconceived ideas 
and to independently confirm the preliminary conclusions of Sec.~\ref{sec:var}.  More specifically, 
we apply to our 2 data sets our now-well-tested tomographic imaging technique, described extensively 
in previous similar studies \citep[e.g.,][]{Donati10b, Donati11}.

Our basic assumption is that the observed profile variations are mainly due to rotational modulation.  
We thus attempt, following the principles of maximum entropy, at simultaneously 
and automatically recovering the simplest magnetic topology, photospheric brightness image and 
accretion-powered \caii\ emission map that are compatible with the series of observed Stokes 
$I$ and $V$ LSD and \caii\ IRT profiles.  For a full description of the imaging method, the 
reader is referred to the previous papers in the series \citep[e.g.,][]{Donati10b, Donati11}.  

\begin{figure*}
\center{\hbox{
\includegraphics[scale=0.37,angle=-90]{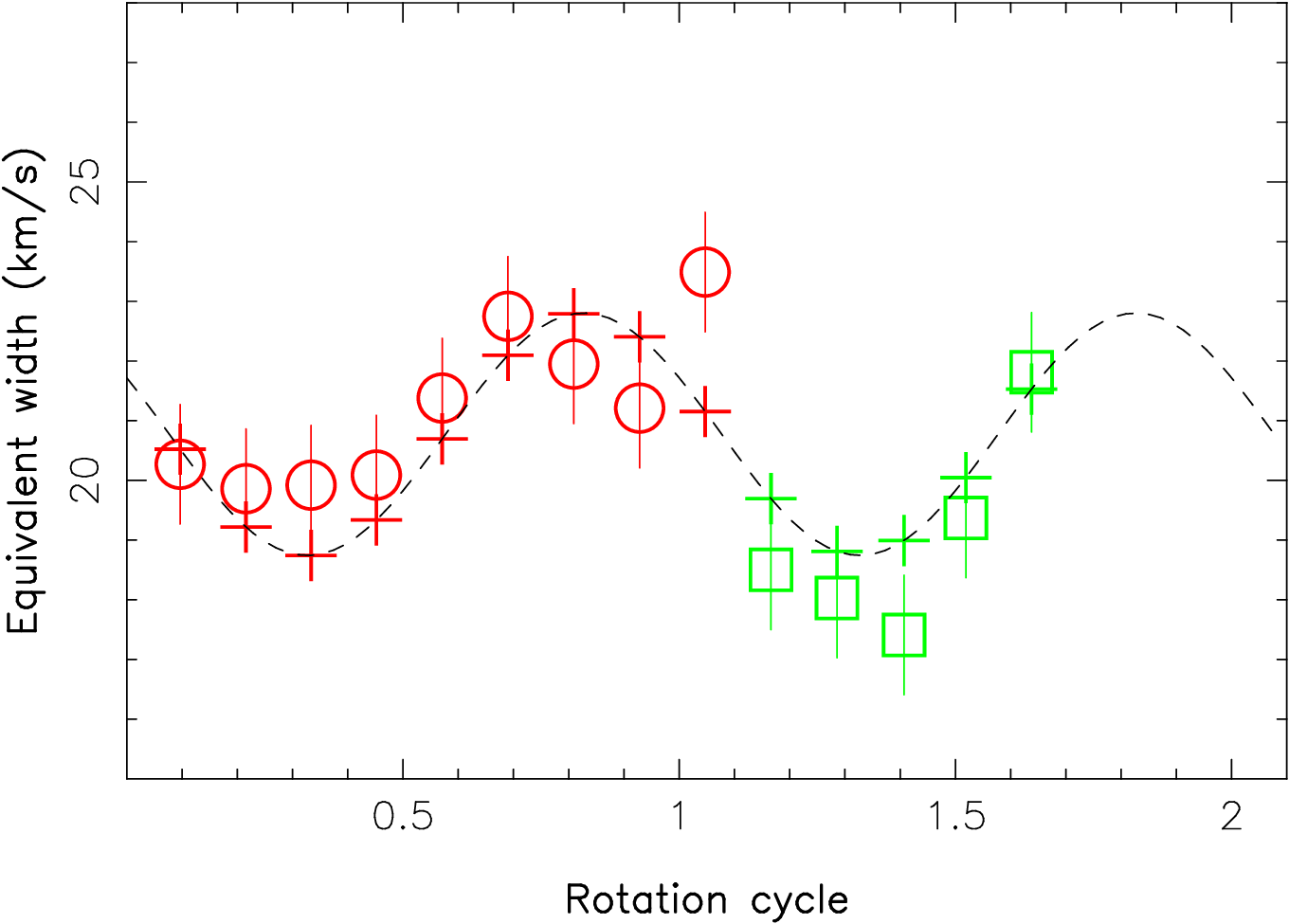}\hspace{5mm}
\includegraphics[scale=0.37,angle=-90]{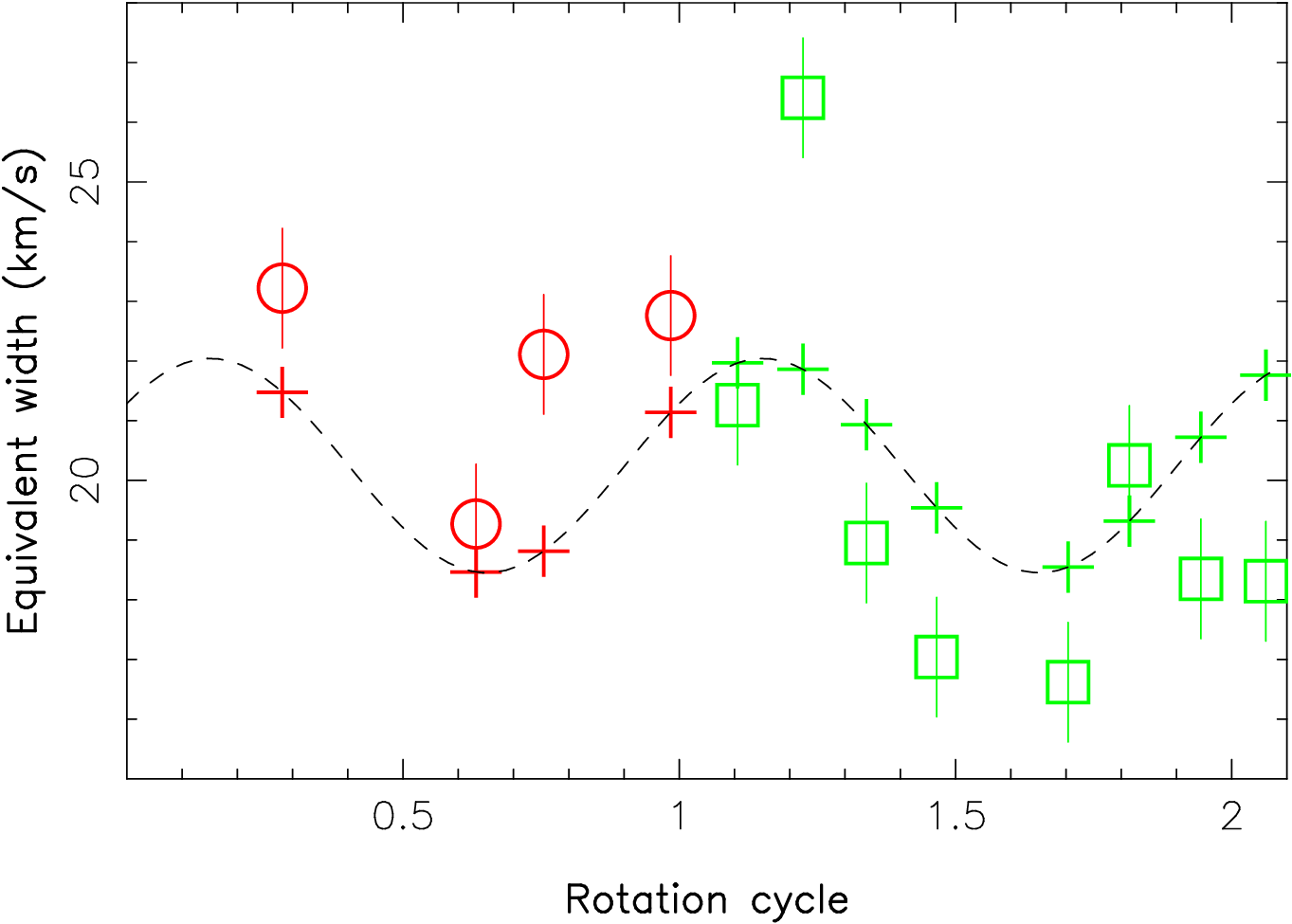}}}
\caption[]{Measured (open symbols) and fitted (pluses) equivalent widths of
the \caii\ IRT LSD profiles of GQ~Lup in 2009~July (left panel) and 2011~June (right panel).
The model wave (dashed line) providing the best (sine+cosine) fit to
the data presumably traces rotational modulation (with a period of 8.4~d), while the deviation 
from the fit illustrates the level of intrinsic variability (significantly higher in 
2011~June).  The open symbols are defined as described in Fig.~\ref{fig:var1}.  }
\label{fig:ew}
\end{figure*}

\subsection{Application to GQ~Lup}

We start by applying to the data the usual filtering procedure \citep[e.g.,][]{Donati10b, Donati11}, 
whose aim is to retain the rotational modulation only and hence help the convergence 
of the imaging code.  
The effect of this filtering on the \caii\ emission flux is shown for instance in Fig.~\ref{fig:ew}.  
We stress that this filtering has little impact on the reconstructed images, results derived from 
the unfiltered data set being virtually identical to those presented below.  

We use Unno-Rachkovsky's equations known to provide a good description of the local Stokes $I$ and
$V$ profiles
(including magneto-optical effects) in the presence of both weak and strong magnetic fields
\citep[e.g.,][Sec.~9.8]{Landi04}.  
The model parameters used for GQ~Lup are mostly identical to those used in our
previous studies \citep{Donati10b, Donati11};  more specifically, the wavelength, Doppler width, 
unveiled equivalent width and Land\'e factor of the average photospheric profile are set to  
640~nm, 1.9~\kms, 4.2~\kms\ and 1.2, while those of the quiet \caii\ profile are set to 
850~nm, 7~\kms, 10~\kms\ and 1.0.  
The emission profile scaling factor $\epsilon$, describing the emission
enhancement of accretion regions over the quiet chromosphere, is once again set to $\epsilon=10$ 
(this choice being somewhat arbitrary as outlined in \citealt{Donati10b}, but not affecting 
significantly the location and shape of features in the reconstructed maps of excess \caii\ emission).  

\begin{figure*}
\vspace{-2mm}
\hbox{\includegraphics[scale=0.7]{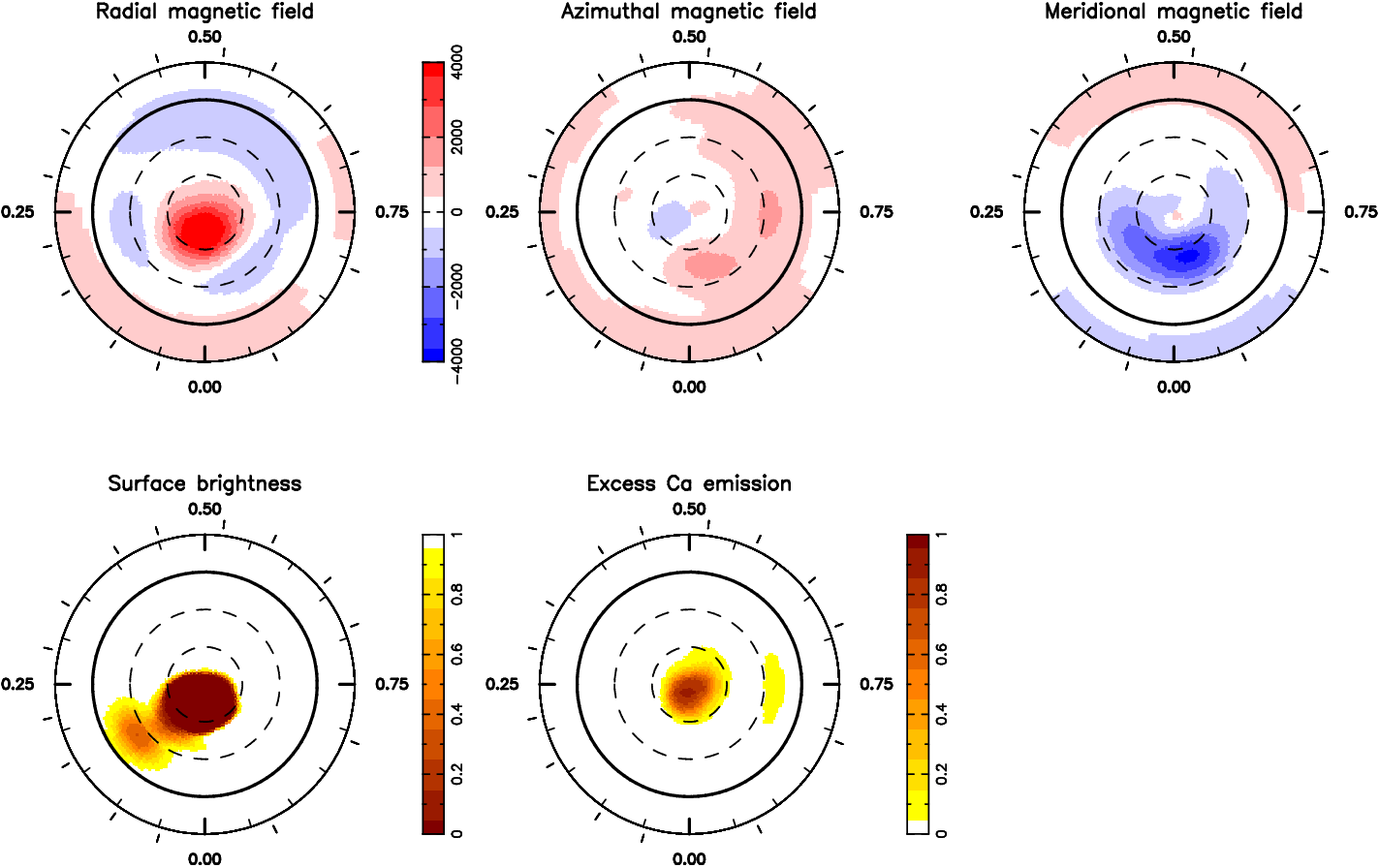}}
\vspace{8mm}
\hbox{\includegraphics[scale=0.7]{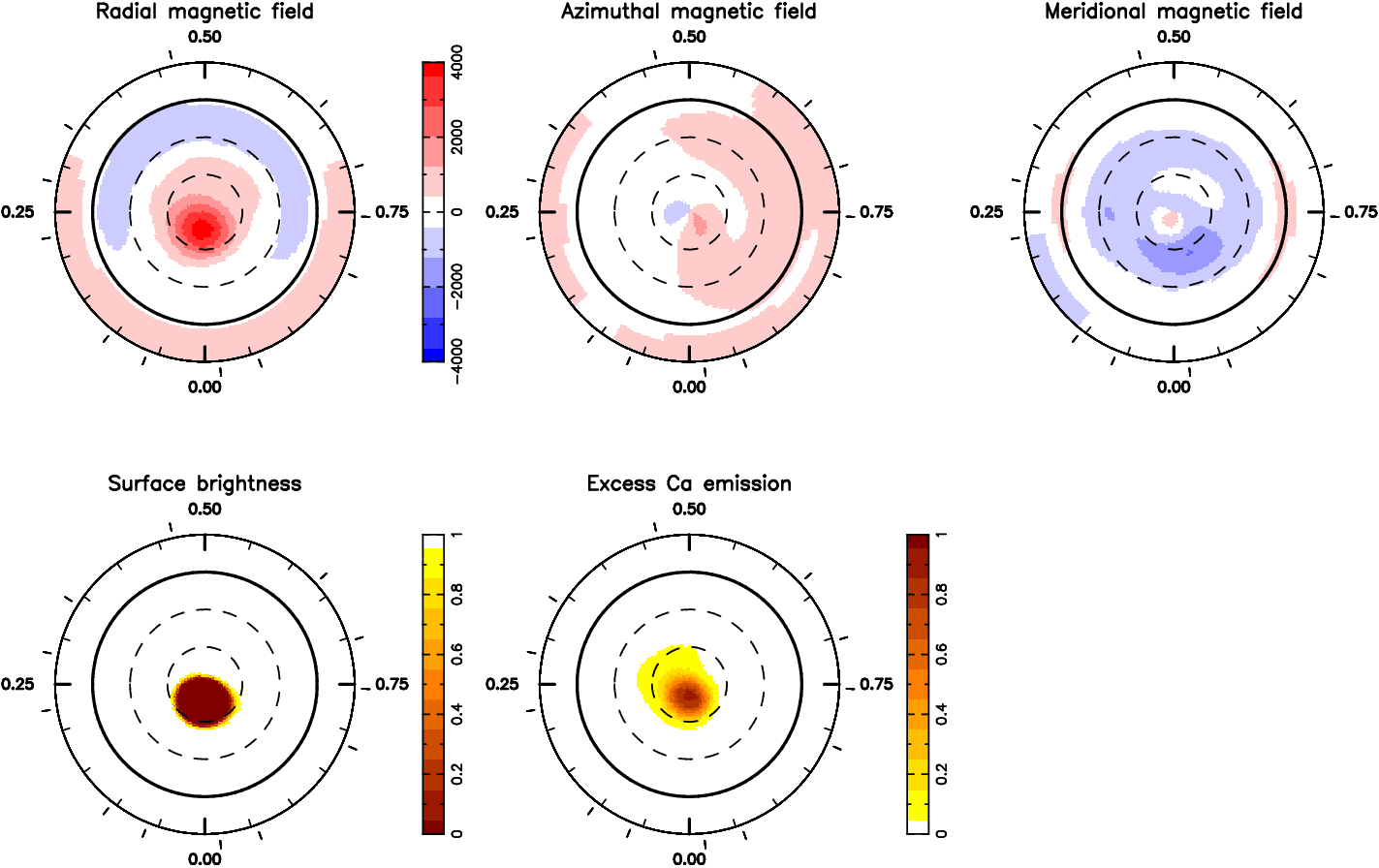}}
\caption[]{Maps of the radial, azimuthal and meridional components of the magnetic field $\bf B$
(first and third rows, left to right panels respectively), photospheric brightness and excess
\caii\ IRT emission (second and fourth rows, first and second panels respectively) at the
surface of GQ~Lup, in 2009~July (top two rows) and 2011~June (bottom two rows).
Magnetic fluxes are labelled in G;  local photospheric brightness (normalized to that of the quiet
photosphere) varies from 1 (no spot) to 0 (no light);  local excess \caii\ emission varies from 0
(no excess emission) to 1 (excess emission covering 100\% of the local grid cell, assuming an
intrinsic excess emission of 10$\times$ the quiet chromospheric emission).
In all panels, the star is shown in flattened polar projection down to latitudes of $-30\degr$,
with the equator depicted as a bold circle and parallels as dashed circles.  Radial ticks around
each plot indicate phases of observations. }
\label{fig:map}
\end{figure*}

\begin{figure*}
\vspace{-3mm}
\center{
\includegraphics[scale=0.65,angle=-90]{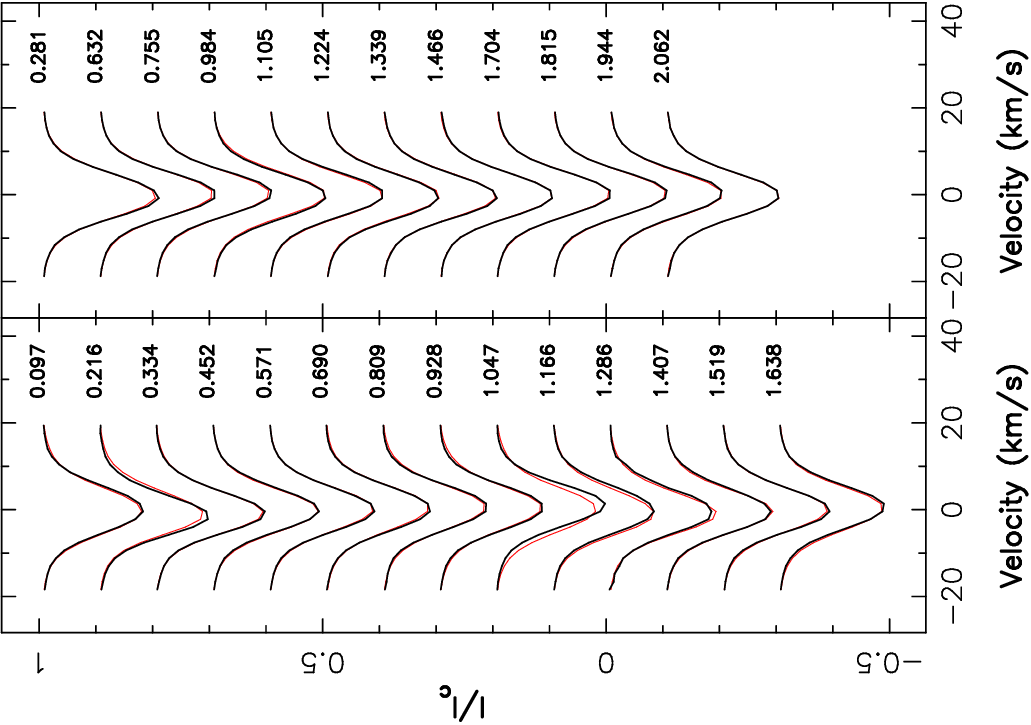}\hspace{4mm}
\includegraphics[scale=0.65,angle=-90]{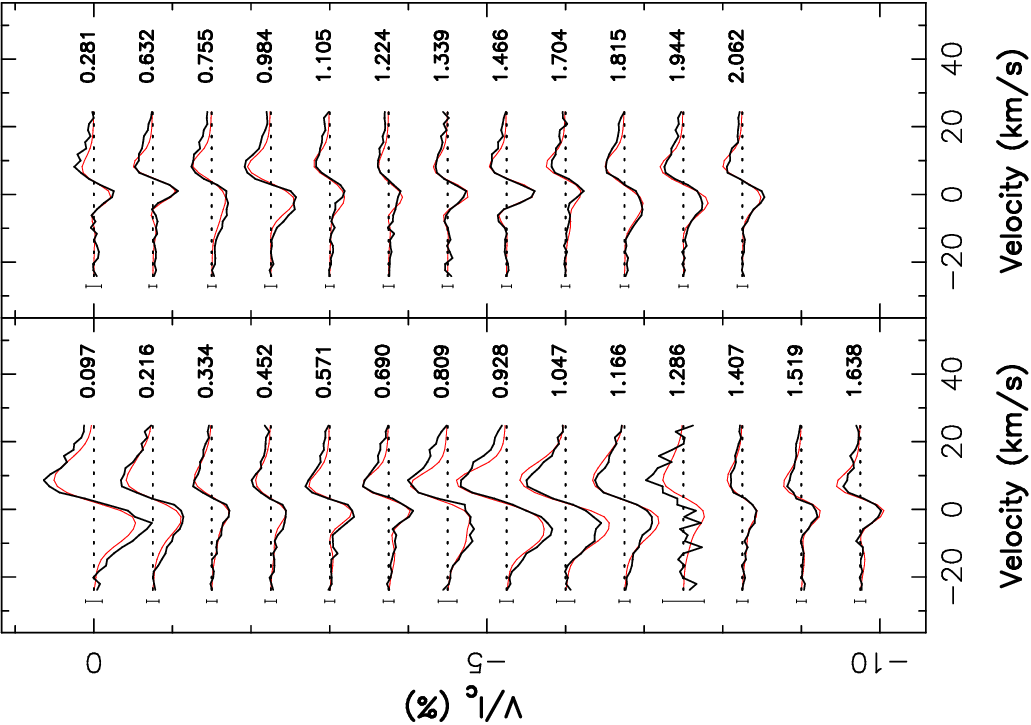}}
\vspace{3mm}
\center{
\includegraphics[scale=0.65,angle=-90]{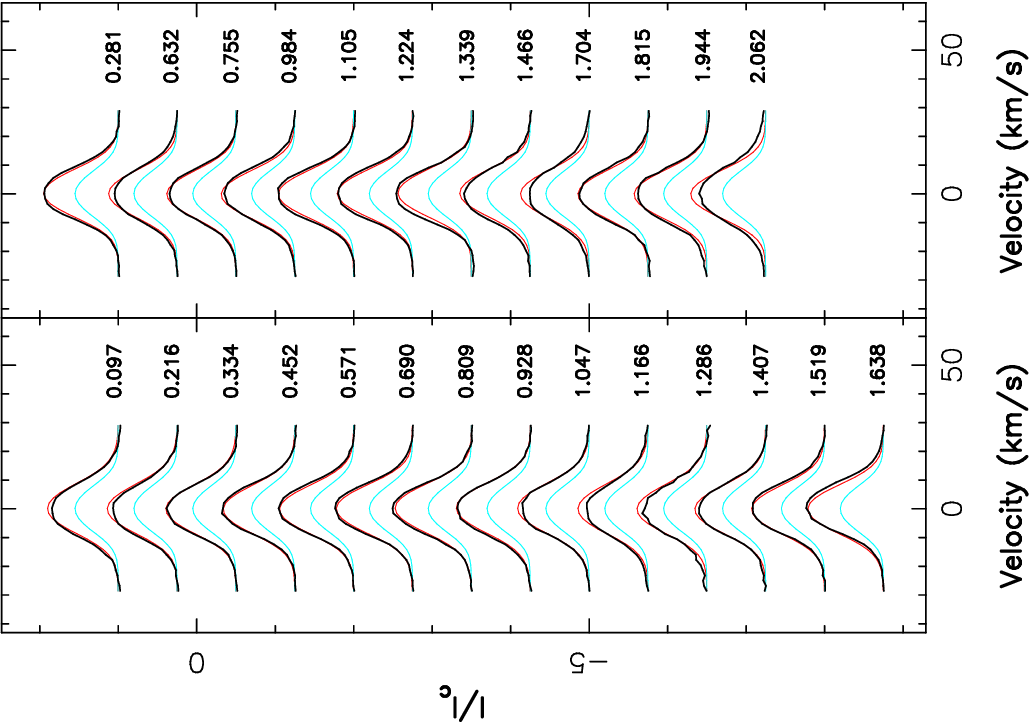}\hspace{4mm}
\includegraphics[scale=0.65,angle=-90]{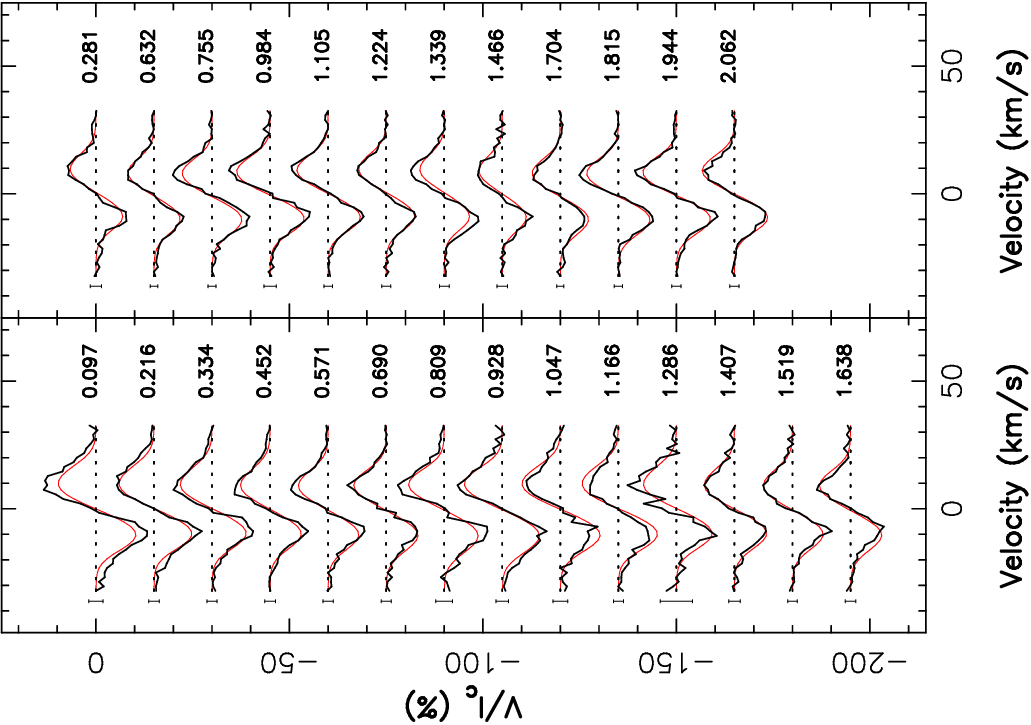}}
\caption[]{Maximum-entropy fit (thin red line) to the observed (thick black line) Stokes $I$ and
Stokes $V$ LSD photospheric profiles (top panels) and \caii\ IRT profiles (bottom panels)
of GQ~Lup.  In each panel, the left and right columns correspond to the 2009~July and 2011~June
data respectively.  The light-blue curve in the bottom left panel shows the (constant)
contribution of the quiet chromosphere to the Stokes $I$ \caii\ profiles.
Rotational cycles and 3$\sigma$ error bars (for Stokes $V$ profiles) are also shown next to each
profile.  }
\label{fig:fit}
\end{figure*}

The magnetic, brightness and accretion maps we reconstruct for GQ~Lup at both epochs are
shown in Fig.~\ref{fig:map}, with the corresponding fits to the data shown in Fig.~\ref{fig:fit}.
The spherical harmonic (SH) expansions describing the field was limited to terms 
with $\ell\leq7$, which is found to be adequate when \vsini\ is low.  
The magnetic topology of GQ~Lup was also assumed to be antisymmetric
with respect to the centre of the star, as in most previous studies.  

Error bars on Zeeman signatures were artificially expanded by a factor of 2 
(both for LSD profiles and for \caii\ emission and at both epochs) to take into account 
the level of intrinsic variability (obvious from the lower panels of Figs.~\ref{fig:var1} and 
\ref{fig:var2}, where the observed dispersion on longitudinal fields is larger than formal 
error bars) and to avoid the code attempting to overfit the data.  
The fits we finally obtain correspond to a reduced chi-square \chisqr\ equal to 1, starting 
from initial values of 16 at both epochs (corresponding to a null magnetic field and unspotted 
brightness and accretion maps, and with scaled-up error bars on Zeeman signatures).

As a by-product, we obtain new estimates for various spectral characteristics of GQ~Lup.
In particular, we find that the average RV of GQ~Lup has changed from --3.2 from --2.8~\kms\ 
between 2009~July and 2011~June;  we believe that this change, about 4 times larger than the 
(conservative) error bar on each measurement, is real.  
We also find that \vsini\ is equal to $5\pm1$~\kms;  this is slightly smaller that the best 
estimate available in the literature \citep[equal to $6.8\pm0.4$~\kms,][]{Guenther05}, though 
presumably more accurate (despite the larger error bar) since magnetic broadening (significant 
in GQ~Lup) is taken into account in our study (and not in the former one).  We find no clear 
evidence that GQ~Lup is not rotating rigidly;  the method previously used for estimating 
differential rotation on V2129~Oph \citep[][]{Donati11} is much less sensitive for GQ~Lup
(given the lower inclination angle and the lower \vsini) and remains inconclusive.  

Finally, we find that the profiles are best fitted for values of the local filling factor 
$\psi$ \citep[describing the relative proportion of magnetic areas at any given point of the 
stellar surface, see][]{Donati10b} equal to $\simeq$0.3 for LSD profiles and $\simeq$0.6 for 
\caii\ emission profiles;  we obtain in particular that using two different $\psi$ values for 
the two sets of lines allows an easier fit to the data, although using the same value for both 
lines \citep[set to $\psi=0.4$, as for all previous cTTSs analysed to date, e.g.,][]{Donati11} 
is still possible (and generates very similar results).  

\subsection{Modelling results}

The large-scale magnetic topology we reconstruct for GQ~Lup is mostly axisymmetric at both 
epochs.  It features in particular a region of positive radial field near the pole (see 
Fig.~\ref{fig:map}), reaching intensities of up to 5.2 and 4.0~kG in 2009~July and 2011~June 
respectively, in rough agreement with the peak longitudinal field values probed by the \hei\ 
emission line (see Fig.~\ref{fig:var3}, lower panels);  it also includes an incomplete ring of 
much weaker negative radial field at low latitudes (see Fig.~\ref{fig:map}).  Both maps 
also include a conspicuous (though irregular) ring of negative (i.e., equatorward) meridional
field located at intermediate latitudes, i.e., between the polar region and low-latitude ring of 
positive and negative radial field respectively.  

The reconstructed field is mostly poloidal, with less that 20\% of the reconstructed magnetic 
energy stored in the toroidal component at either epochs.  The poloidal component is mainly 
axisymmetric, concentrating at least 80--85\% of the magnetic energy in SH modes with 
$m<\ell/2$.  More specifically, the poloidal field is found to be dominantly octupolar 
at both epochs, 
as obvious from both radial and meridional field maps;  the octupole is aligned with the 
rotation axis to within better than 10\degr\ and reaches 
strengths of 2.4 and 1.6~kG in 2009~July and 2011~June respectively.  The large-scale dipole 
is typically half as intense as the large-scale octupole, with strengths of 1.1 and 0.9~kG in 
2009~July and 2011~June (implying octupole to dipole polar strength ratios of 2.2 and 1.8 
respectively);  it is also more or less parallel (rather than anti-parallel) to the 
octupole, and is tilted by about 30\degr\ to the rotation axis (towards phase 0.05) at both 
epochs.  We can therefore safely confirm that the large-scale field of GQ~Lup significantly 
weakened between 2009~July and 2011~June, as guessed from the long-term evolution of the 
longitudinal field curves (see Sec.~\ref{sec:var});  we can further conclude that this 
variation is mostly attributable to a weakening (by about 1/3) of the octupole component, 
the overall topology of GQ~Lup remaining dominantly octupolar at both epochs.  

The reconstructed photospheric brightness distribution mostly features a cool dark spot 
near the pole (see
Fig.~\ref{fig:map}), shifted by about 20\degr\ towards phase 0.05 and occupying about 8\% and 
4\% of the total stellar surface in 2009~July and 2011~June respectively.  This cool polar 
spot mostly overlaps with the visible magnetic pole, as in most other cTTSs magnetically 
imaged to date;  it confirms a posteriori that this is the reason why LSD profiles of 
photospheric lines are probing field regions of opposite polarity than the accretion lines, 
i.e., the weakly magnetic low-latitude regions (of negative radial field polarity) rather 
than the strongly magnetic polar regions (of positive radial field polarity) which emit 
very few photons per unit area (relative to the rest of the star) and therefore contribute 
very little to both Stokes $I$ and $V$ LSD profiles.  
The cool polar spot is found to extend to low latitudes in 2009~July, as expected from the 
larger amplitude of the RV curve from photospheric lines (see Sec.~\ref{sec:var} and 
Fig.~\ref{fig:var1}, upper panels).  We note that synthetic phased curves of RVs and 
longitudinal fields derived from the reconstructed magnetic and brightness maps are in 
good agreement with observations at both epochs.  

The maps of excess \caii\ emission also show a clear accretion region close to the pole 
(see Fig.~\ref{fig:map}), slightly shifted towards phase 0.0 by about 10\degr\ 
and covering $\simeq$2.5\% of the stellar surface.  We note that, in 2009~July, the accretion 
map also feature a low-contrast crescent-shape region at low latitudes, centred at phase 0.75; 
apart from this small difference, the two maps are rather similar, which further demonstrates, 
at the same time, that the change we report in the longitudinal field curves of accretion 
proxies (see Sec.~\ref{sec:var}) unambiguously traces a temporal evolution of the large-scale 
magnetic topology (and not of the accretion pattern).  Unsurprisingly, the synthetic RV curve 
derived from the 2009~July accretion map, reaching maximum and minimum at phases 0.2 
and 0.6 respectively, is found to be more or less in anti-phase with the corresponding 
observations (see top-left panel of Fig.~\ref{fig:var2}) though both are showing a similarly low
peak-to-peak amplitude (of $\simeq$0.5~\kms);  
this confirms the unusual (and yet unexplained) behaviour of the \caii\ RVs at this epoch  
that we already mentioned in Sec.~\ref{sec:var}.  All other synthetic phased curves (of either 
equivalent widths, RVs and longitudinal fields) are otherwise in good agreement with 
observations at both epochs. 


\section{Summary \& discussion}
\label{sec:dis}

This paper presents the first spectropolarimetric analysis of the cTTS GQ~Lup, following 
previous similar studies of several cTTSs of various masses and ages;  this analysis uses 
extensive data sets collected at two different epochs (2009~July and 2011~June), in the 
framework of the MaPP Large Program with ESPaDOnS at CFHT.  From these data, we start by 
redetermining the fundamental characteristics of GQ~Lup, and in particular its 
photospheric temperature, found to be $\simeq$250~K warmer than usually quoted in the literature.  
We also obtain that the rotation period of GQ~Lup is $8.4\pm0.3$~d, in good agreement 
with the previous estimate of \citet{Broeg07}.  
We finally conclude that GQ~Lup is a $1.05\pm0.07$~\msun\ star with an age of 2--5~Myr and a 
radius of $1.7\pm0.2$~\rsun\ that has just started to build a radiative core (see 
Fig~\ref{fig:hrd}).  

Strong Zeeman signatures are detected at all times in the spectra of GQ~Lup, both in LSD 
profiles of photospheric lines and in emission lines probing accretion regions at the 
chromospheric level.  We report longitudinal fields ranging from --0.1 to --0.6~kG in 
photospheric lines,  from 0.7 to 1.9~kG in the emission core of \caii\ lines, and from 
1 to more than 6~kG in the narrow emission profile of \hei\ $D_3$ lines, making GQ~Lup 
the cTTS with strongest magnetic fields known as of today.  
We find in particular that different field polarities are traced by photospheric lines and 
accretion proxies (as for several other cTTSs), indicating that they probe different 
spatial regions of GQ~Lup.  Longitudinal field curves also unambiguously demonstrate that 
the parent large-scale field of GQ~Lup significantly evolved between our two observing 
runs, with magnetic intensities in accretion regions (where the field is strongest) 
dropping by as much as 50\% between 2009~July and 2011~June;  this makes GQ~Lup the 
second cTTS on which temporal evolution of the large-scale magnetic topology has 
been unambiguously demonstrated \citep[after V2129~Oph,][]{Donati11}.  

Using our tomographic imaging tool specifically adapted to the case of cTTSs, we 
convert our two data sets into surface maps of GQ~Lup, of the large-scale vector magnetic 
field on the one hand, and of the photospheric brightness and of the accretion-powered 
excess emission on the other hand.  We find in particular that the large-scale field of 
GQ~Lup is strong, and mostly poloidal and axisymmetric (about the rotation axis).  
More specifically, we find that the poloidal field is dominated by an octupolar component 
aligned with the rotation axis (within 10\degr) and whose strength weakens from 2.4~kG 
in 2009~July to 1.6~kG in 2011~June;  we also find that the large-scale dipole of GQ~Lup 
is about half as strong as the octupole, with a polar strength equal to 1.1 and 0.9~kG 
in 2009 and 2011, and is tilted by $\simeq$30\degr\ to the rotation axis (and thus largely 
parallel, rather than anti-parallel, to the octupole).  

The large-scale magnetic topology that we reconstruct for GQ~Lup is fully compatible with 
previous results obtained on cTTSs, suggesting that large-scale fields of protostars 
hosting small radiative cores are mostly poloidal and axisymmetric, with a dominant 
octupolar component \citep[e.g.,][]{Donati11}.  This is also similar to that found on 
main-sequence M dwarfs, where stars with radiative cores smaller than 0.5~\rstar\ host 
rather strong and mainly poloidal and axisymmetric fields \citep[e.g.,][Gregory et al.\ 2012, submitted]{Morin08b}.  
This new result further argues that the magnetic fields of cTTSs are generated through dynamo 
processes rather than being fossil fields initially present in the parent cloud from 
which the star has formed and that unexpectedly managed to survive the various turbulent 
episodes of the cloud-to-disc and disc-to-star contraction phases;  clear observational 
evidence that the large-scale field significantly weakened in a 2~yr timescale 
independently confirms this conclusion, and indicates at the same time that the underlying 
dynamo processes are non-stationary.  

We also find that the visible pole of GQ~Lup (and presumably the invisible pole as well) hosts 
a cool dark spot at photospheric level and concentrates most of the accreted material from 
the disc, if we judge from the location of the accretion-powered area of excess \caii\ emission 
at chromospheric level (roughly overlapping with the cool photospheric spot).  These near-polar 
regions are consistent with the observed low-amplitude RV rotational modulation of photospheric 
lines and accretion proxies.  We also note that additional low-latitude features are detected 
in 2009~July, both in the photospheric brightness map and in the distribution of 
accretion-powered emission, suggesting that accretion may also occur (though marginally) 
at low latitudes at this epoch;  a similar conclusion was proposed by \citet{Broeg07} to 
attempt reconciling the low inclination of GQ~Lup with the large amplitude of its light curve 
at some epochs.  As proposed below, this may be related to the stronger octupolar component 
of the large-scale magnetic field in the first of our two observing runs.  

Given the emission fluxes of the conventional accretion proxies, we infer that the logarithmic 
mass accretion rate at the surface of GQ~Lup is equal to $-9.0\pm0.3$ (in \mspy), being 
slightly stronger (by $\simeq$0.2~dex) in 2009~July than in 2011~June.  
From this, we infer that GQ~Lup should be capable of magnetically disrupting its accretion 
disc up to a radius of $\rmag=4.6\pm1.0$~\rstar, or equivalently $0.037\pm0.008$~au 
\citep[assuming an average dipole strength of 1~kG and using the analytical formula 
of][]{Bessolaz08};  in particular, we find that \rmag\ is significantly larger than the 
radius at which the dipole field starts to dominate over the octupole field \citep[equal to 
$\simeq$1.2~\rstar, for an octupole about twice stronger than the dipole,][]{Gregory11}, 
in agreement with our observation that the accretion flow is mostly poleward on GQ~Lup.  
For larger octupole to dipole polar strength ratios (and when the dipole and octupole are 
parallel rather than antiparallel), one would naturally expect GQ~Lup to trigger increasingly 
larger amounts of low-latitude accretion \citep[e.g.,][]{Romanova11};  this may be what is 
happening to GQ~Lup in 2009~July, i.e., when the octupole component is strongest and low-contrast 
low-latitude accretion signatures are being detected.  More observations are however necessary to 
confirm whether observations are consistent with theoretical expectations on this specific issue.  

When \rmag\ is compared to the corotation radius $\rcor\simeq10.4$~\rstar\ (or 0.083~au), at which the 
Keplerian period equals the stellar rotation period, we find that $\rmag/\rcor$ is equal to 
$0.45\pm0.10$, far below the value (of $\simeq1$) at which star/disc magnetic coupling can start 
inducing a significant outflow through a propeller-like mechanism \citep[e.g.,][]{Romanova04} 
and thus force the star to spin down.  We thus speculate that, following the recent build-up 
of a radiative core and the corresponding change in its large-scale magnetic topology, 
GQ~Lup can no longer counteract the increase of its angular momentum (resulting from both 
contraction and accretion) through a star-disc braking torque, and has no other option than 
to undergo a rapid spin up such as that TW~Hya experienced already \citep{Donati11b}.  
Again, more observations are needed to check these speculations, and in particular to 
monitor how the magnetic field of the roughly solar-mass GQ~Lup is evolving on a longer term 
(at least on a timescale comparable to that of the solar cycle) and to validate whether the 
topology we reconstructed from our 2009 and 2011 data sets is indeed typical and adequate to 
predict the rotational evolution of GQ~Lup.  

We complete this section by briefly discussing the RV change that we report for GQ~Lup, 
whose average LSD photospheric profile was observed to shift from $-3.2\pm0.1$~\kms\ in 
2009~July to $-2.8\pm0.1$~\kms\ in 2011~June, much larger in particular than spurious shifts 
potentially attributable to instrument stability problems.  Being comparable to the 
semi-amplitude of the activity-induced modulation of the RV curve (and readily visible from 
the plots themselves, see Fig.~\ref{fig:var1}, upper panels), this change is also undoubtedly 
too large to be realistically attributed to uncorrected residuals of RV effects induced by 
cool spot patterns at the surface of GQ~Lup.  
Various possibilities thus remain to explain it.  The first 
option is that it is caused by long-term changes of the surface convection / granulation 
pattern;  although 0.4~\kms\ may sound fairly large (for a Sun-like star), little is known 
in practice about how much this effect can modify RVs in stars as active as cTTSs.  
The second option is that it is caused by an additional body orbiting GQ~Lup, on a timescale 
of months or years.  We already know that GQ~Lup is orbited by a brown dwarf \citep{Neuhauser05, 
McElwain07, Lavigne09}, but its 
distance from GQ~Lup (about 100~au) is likely too large to induce a RV shift similar to 
the one we report here in a timescale of only 2~yr;  this may suggest that GQ~Lup hosts a 
third body much closer to the central protostar than GQ~Lup~B, e.g., a brown dwarf of a few 
tens of Jupiter masses at a distance of a few au's.  Additional observations of GQ~Lup similar 
to those presented here are obviously needed to confirm which option applies.  

\section*{Acknowledgements}
We thank the anonymous referee for valuable comments and suggestions that improved the overall 
clarity of the manuscript.  This paper is based on
observations obtained at the Canada-France-Hawaii Telescope (CFHT), operated by the National
Research Council of Canada, the Institut National des Sciences de l'Univers of the Centre
National de la Recherche Scientifique of France and the University of Hawaii.
The ``Magnetic Protostars and Planets'' (MaPP) project is supported by the
funding agencies of CFHT and TBL (through the allocation of telescope time)
and by CNRS/INSU in particular, as well as by the French ``Agence Nationale
pour la Recherche'' (ANR).
SGG is supported by NASA grant HST-GO-11616.07-A;  SHPA acknowledges support from CNPq, CAPES and Fapemig.

\bibliography{gqlup}
\bibliographystyle{mn2e}
\end{document}